\documentclass[a4paper,12pt]{article}
\pdfoutput=1
\usepackage{a4wide}
\usepackage{amsmath,amsfonts,amssymb,cite}

\usepackage{bm}
\usepackage{subfigure}

\usepackage{graphicx}
\usepackage{enumerate}
\usepackage{epsfig}
\usepackage{soul}
\usepackage{verbatim,setspace}
\usepackage{tikz}
\usetikzlibrary{calc}
\newcommand{\cL}{{\cal L}}

\newcommand{\ov}{\overline}

\newcommand{\be}{\begin{equation}}
\newcommand{\ee}{\end{equation}}
\newcommand{\bea}{\begin{eqnarray}}
\newcommand{\eea}{\end{eqnarray}}

\newcommand{\MET}{E\llap{/\kern1.5pt}_T}
\newcommand{\nn}{\nonumber}

\def\nn{\nonumber}

\def\beq{\begin{equation}}
\def\eeq{\end{equation}}

\def\psqr#1#2{{\vcenter{\vbox{\hrule height.#2pt
        \hbox{\vrule width.#2pt height#1pt \kern#1pt
        \vrule width.#2pt}
        \hrule height.#2pt \hrule height.#2pt
        \hbox{\vrule width.#2pt height#1pt \kern#1pt
        \vrule width.#2pt}
        \hrule height.#2pt}}}}
\def\sqr#1#2{{\vcenter{\vbox{\hrule height.#2pt
        \hbox{\vrule width.#2pt height#1pt \kern#1pt
        \vrule width.#2pt}
        \hrule height.#2pt}}}}

\def\ov{\overline}

\newcounter{oldcounter}
\addtocounter{equation}{1}
\begin{document}

\vspace*{-2cm}
\begin{flushright}
DESY-15-231
\end{flushright}

\begin{center}
{\Large \bf Signatures of top flavour-changing dark matter}\\
\vspace{0.3cm}
{\bf Jorgen D'Hondt$^a$, Alberto Mariotti$^a$, Kentarou Mawatari$^{a,b}$,\\
 Seth Moortgat$^a$, Pantelis Tziveloglou$^a$, Gerrit Van Onsem$^{a,c}$}
\vspace{0.3cm}

\noindent \textit{{\small 
$^a$\,Theoretische Natuurkunde and IIHE/ELEM, Vrije Universiteit Brussel\\
International Solvay Institutes \\ 
Pleinlaan 2, B-1050 Brussels, Belgium \\[2mm]
$^{b}$\,Laboratoire de Physique Subatomique et de Cosmologie, Universit\'e
Grenoble-Alpes,\\
CNRS/IN2P3, 
53 Avenue des Martyrs, F-38026 Grenoble, France \\[2mm]
$^{c}$\,DESY, Notkestr. 85, D-22607 Hamburg, Germany
}}
\end{center}

\vspace{1.5cm}
\begin{abstract}
\noindent \normalsize
We develop the phenomenology of scenarios in which a dark matter
 candidate interacts with a top quark through flavour-changing
 couplings, employing a simplified dark matter model with an $s$-channel
 vector-like mediator.   
We study in detail the top--charm flavour-changing interaction, by
 investigating the single top plus large missing energy signature at the
 LHC as well as constraints from the relic density and direct and
 indirect dark matter detection experiments. 
We present strategies
 to distinguish between the top--charm and top--up flavour-changing
 models by taking advantage of the lepton charge asymmetry as well as by
 using charm-tagging techniques on an extra jet.
We also show the complementarity between the LHC and canonical dark
 matter experiments in exploring the viable parameter space of the
 models. 
\end{abstract}

\newpage

\tableofcontents

%\setcounter{page}{1}
%\twocolumn[]

%%%%%%%%%%%%%
\section{Introduction}
\label{intro}
%%%%%%%%%%%%%

The description of dark matter (DM), whose abundant presence in the universe is supported by overwhelming observational evidence, is nowadays one of the main motivations for physics beyond the Standard Model (SM). The usual paradigm to realize the correct relic abundance of DM in the
universe relies on a weak coupling between the SM particles and the DM candidate, the so-called weakly interacting massive particle (WIMP). This scenario implies that signatures of DM could be discovered in colliders and canonical DM experiments, such as direct and indirect detection experiments. Indeed, in the last years there have been intense activities in developing DM searches at the LHC and their possible interplay with underground and satellite DM experiments.

Since the properties and the interaction of DM with SM particles are unknown, a current bottom-up approach is to employ simplified models of DM that capture the phenomenology of different types of theories beyond the SM. Simplified models typically consist of new DM species and mediator fields that connect the SM and the DM sectors. They are usually parametrised by the masses of the mediator and DM fields as well as the couplings of the mediator to the DM and the SM particles (see e.g.~\cite{Abercrombie:2015gea}).

One interesting possibility among various types of simplified models is that the DM sector couples to the SM through 
flavour non-universal interactions. In this context, one can envisage
simplified DM models with flavour violating structures and identify the relevant constraints as well as the possible novel signatures of such scenarios. Non-universal flavour couplings can arise in $Z'$
models~\cite{Jung:2009jz,Andrea:2011ws,Boucheneb:2014wza} and in
the flavoured DM paradigm~\cite{Kumar:2013hfa,Lopez-Honorez:2013wla,Kile:2011mn,Batell:2011tc,Kamenik:2011nb,Agrawal:2011ze,Agrawal:2014una,Agrawal:2014aoa,Calibbi:2015sfa,Bishara:2015mha}, where the DM candidate belongs to a sector which is also responsible to
explain the flavour structure in the SM. 

Among the non-universal flavour interactions, the couplings that involve third generation quarks are less constrained by low-energy experiments. Moreover, the top quark, being the heaviest of the SM particles, 
can be an interesting candidate to represent the portal through which DM couples to
the SM sector.  
The LHC signatures and the DM constraints on such models are in general different from those of the usual WIMP models. 
Flavour non-universal yet diagonal DM models characterised by the
four-fermion interaction between a DM pair and a top-quark pair have
been studied in an effective field theory (EFT)  
approach~\cite{Cheung:2010zf,Haisch:2012kf,Lin:2013sca} and searched for
in a top-quark pair plus missing energy final state at the LHC
Run-I~\cite{Aad:2014vea,Khachatryan:2015nua}.    
Recently more studies in simplified models involving top quarks have
been 
done~\cite{Buckley:2014fba,Haisch:2015ioa,Mattelaer:2015haa,Backovic:2015soa}.

It is interesting to extend these studies to a possibility of having a
flavour-changing coupling involving the top quarks.  
Inclusion of the top flavour-changing interaction with DM opens a peculiar
collider signature, the so-called
monotop signature~\cite{Andrea:2011ws,Kamenik:2011nb,Fuks:2014lva}, already
searched for by the ATLAS and CMS
Collaborations~\cite{Aad:2014wza,Khachatryan:2014uma}. 
We note that in the past the top--up flavour-changing coupling has been 
extensively
studied~\cite{Andrea:2011ws,Kamenik:2011nb,Wang:2011uxa,Kumar:2013jgb,Alvarez:2013jqa,Agram:2013wda,Ng:2014pqa,Boucheneb:2014wza,Fuks:2014lva,Allahverdi:2015mha},
while the top--charm coupling has received less attention.

Moreover, top flavour-changing DM models are interesting also because they can account \cite{Rajaraman:2015xka} 
for the 
excess of gamma rays originating in the centre of our galaxy \cite{Hooper:2010mq,murgia,Porter:2015uaa},
without any conflict with constraints from flavour physics.
We note that there has been recently some debate about the DM origin of this excess and several alternative explanations have been put forward
(see e.g. \cite{Bartels:2015aea,Lee:2015fea,Gaggero:2015nsa,Abazajian:2015raa} and references therein).
Nevertheless it is interesting, during the analysis of our simplified model, to discuss the tantalizing possibility that this is indeed due to a DM signal.

In this work, we study in detail the phenomenology of a simplified DM model with top--charm flavour-changing
interactions, highlighting the difference from the top--up flavour-changing case.
We present the prospects for the LHC Run-II, concentrating on the
leptonic single-top final state. We also investigate the charge asymmetry of the lepton in the final state and
charm-tagging techniques to
distinguish between the top--charm and top--up interaction models.  
Apart from the LHC DM searches, we discuss in detail the relic density, and
indirect and direct detection constraints on the model. 
The interplay for DM searches between the LHC and non-collider experiments will reveal different features between
the top--charm flavour-changing interaction and the top--up one, 
furnishing another interesting manifestation of complementarity among different DM search experiments.

The paper is organized as follows.
In Sec.~\ref{sec:models} we introduce the simplified DM model
involving the top flavour-changing interaction, and classify the
signatures depending on the model parameters. 
We also discuss the relation with the EFT approach.
In Sec.~\ref{sec:signatures} we discuss signatures of our model at
the LHC as well as in the relic density, direct and indirect detection
experiments.
Section~\ref{sec:summary} is devoted to our summary and discussion.

%%%%%%%%%%%%%%%%%%
\section{Models}\label{sec:models}
%%%%%%%%%%%%%%%%%%

We start this section by describing simplified models for top
flavour-changing DM and discussing the possible signatures. 
Then, we give some remarks about the similarities and differences of the
signatures between the top--up and top--charm flavour-changing models.
Lastly, we introduce the corresponding EFT description to briefly
mention the relation with the simplified model and its
validity.

\subsection{Simplified models}\label{sec:modelsig}

Since the dynamics of DM are not known, simplified models have been
proposed that parametrise the way DM interacts with SM particles.
In the simplest versions, the SM is extended by two new species, a DM particle
and a particle that mediates the interaction between the DM and SM
particles, called the ``mediator''.  
In this work we are interested in the phenomenological implications of
the interactions of a fermionic Dirac DM ($\chi$) with the
quark sector through an $s$-channel vector-like mediator ($Z'$). 
The interaction Lagrangian is given by
\begin{align}
 \mathcal{L}_{int} =
  g_{\chi}\, \bar \chi \gamma_{\mu} \chi Z'^{\mu} 
  +( g^{Q}_{ij}\, \bar Q_L^i \gamma_{\mu} Q_L^j\,Z'^{\mu} 
    +g^{u}_{ij}\, \bar u_R^i \gamma_{\mu} u_R^j\,Z'^{\mu} 
    +g^{d}_{ij}\, \bar d_R^i \gamma_{\mu} d_R^j\,Z'^{\mu} +h.c.)\,.
\label{Lint}
\end{align}
As mentioned in Sec.~\ref{intro}, we are interested in flavour-changing DM
interactions in the up-quark sector, specifically involving the top
quark.  
If the interactions involve the $SU(2)_L$ doublets, large flavour
off-diagonal couplings in the left-handed sector would imply large
flavour violation also for the down sector, which are strongly
constrained by flavour physics, e.g. by $B_d$--$\bar B_d$
mixing~\cite{Cao:2010zb}. 
Instead, flavour-changing operators involving right-handed top quarks
and up or charm quarks are phenomenologically viable. 
Therefore, we focus on studying the effective flavour-changing
interaction with right-handed up-type quarks in the
Lagrangian~\eqref{Lint}, i.e.   
\begin{align}
 \cL_{int}=g_{\chi}\,\bar{\chi}\gamma^{\mu}\chi Z'^{\mu}
  + ( g_{13}\,\ov{u}_R\gamma_{\mu}t_R\,Z'^{\mu} 
     +g_{23}\,\ov{c}_R\gamma_{\mu}t_R\,Z'^{\mu} +h.c.)\,.
\label{ZprimeLag}
\end{align}
Hereafter we omit the superscript `$u$' of the coupling parameters
$g_{ij}^u$.  
Note that if both the up and charm flavour-changing operators are
present we expect a relevant box diagram contribution to the
$D_0$--$\bar D_0$ mixing~\cite{Ciuchini:2007cw}.   
Hence in the following we consider one of these operators at a time.
\begin{table}
\center
\begin{tabular}{llccccc}
\hline
 && DM annihilation & $\Delta\Gamma_t$ & $\Gamma_{Z'}$ & monotop \\
\hline
 $2m_\chi>m_t$ & $m_{Z'}>2m_\chi>m_t$ & yes & 0 & $tq,\,\chi\chi$ & yes \\
            & $m_{Z'}\sim 2m_\chi>m_t$ & enhanced & 0 & $tq$ & suppressed \\
            & $2m_\chi>m_{Z'}>m_t$ & yes & 0 & $tq$ & suppressed \\ 
 \\
            & $2m_\chi>m_t>m_{Z'}$ & yes & $qZ'$ & suppressed & suppressed \\
\hline
 $m_t>2m_\chi$ & $m_{Z'}>m_t>2m_\chi$ & suppressed & rare $q\chi\chi$ & $tq,\,\chi\chi$ & yes \\
 \\
            & $m_t>m_{Z'}>2m_\chi$ & suppressed & $qZ'$ & $\chi\chi$ & yes \\
            & $m_t>2m_\chi>m_{Z'}$ & suppressed & $qZ'$ & suppressed &
 suppressed \\
\hline
\end{tabular}
\caption{\label{tab:sig}
 Signatures of the top flavour-changing DM model in each mass spectrum. 
 $\Delta\Gamma_t$ and $\Gamma_{Z'}$ represent the partial decay modes of
 the top quark and the $Z'$, respectively.
} 
\end{table}
With the above simplification, the model has in total four parameters,
i.e. two couplings and two masses:
\begin{align}
 \{g_{\chi},\,g_{i3},\,m_{\chi},\,m_{Z'}\}\quad {\rm with}\quad
 i=1\ {\rm or}\ 2\,.
\end{align}
Table~\ref{tab:sig} summarises the various expected signatures in
different regions of the parameter space, which will be discussed
below. 

We note here that, since we intend to focus on the experimental signatures of the top flavour-changing DM model, we postpone to future investigation a detailed study of the possible UV completions.  
At the end of the paper we discuss the basic guidelines and the most relevant issues in constructing such complete theories, identifying possible parallel implications for low-energy phenomenology.

\paragraph{DM annihilation:}

The annihilation of the DM candidate to SM particles during the early universe determines its thermal relic density while late time annihilation in the centre of galaxies offers a possibility to detect DM indirectly, via observation of its SM products. The annihilation channel is 
\begin{align}
 \chi + \bar\chi \to Z'^{(*)} \to t +\bar q \ \ {\rm and}\ \
 \bar t + q\,,
\label{annihilation}
\end{align}
where $q$ denotes an up quark or a charm quark. The annihilation is kinematically efficient for $2m_\chi>m_t$ and is enhanced by threshold effects for $m_{Z'}\sim 2m_\chi$. 
If $m_t>2m_\chi$, the annihilation is suppressed since it occurs via the
off-shell top quark (and/or $W$ boson). Therefore, DM would be overly produced in the early universe, inconsistent with the current observation of the relic abundance of DM. The relic density and the phenomenology of indirect detection experiments are detailed in Sec.~\ref{sec:cDMsearch}.

\paragraph{DM--nucleus scattering:}

Another process that is potentially relevant for the phenomenology of our DM models is the elastic scattering
\begin{align}
 \chi+N\to\chi+N\,,
\end{align}
of DM particles off nuclei of direct detection experiments. Detection of
DM in this type of experiments is based on the observation of the
nuclear recoil energy that the scattering releases. Different from usual
flavour-conserving DM models, the above interaction in our model occurs
only through loop diagrams, and hence it is expected to be strongly
suppressed. In Sec.~\ref{sec:cDMsearch} we discuss further the direct
detection phenomenology and calculate the one-loop process in our model.

\paragraph{Top decay:}

\begin{figure}
\center
 \includegraphics[width=0.95\textwidth]{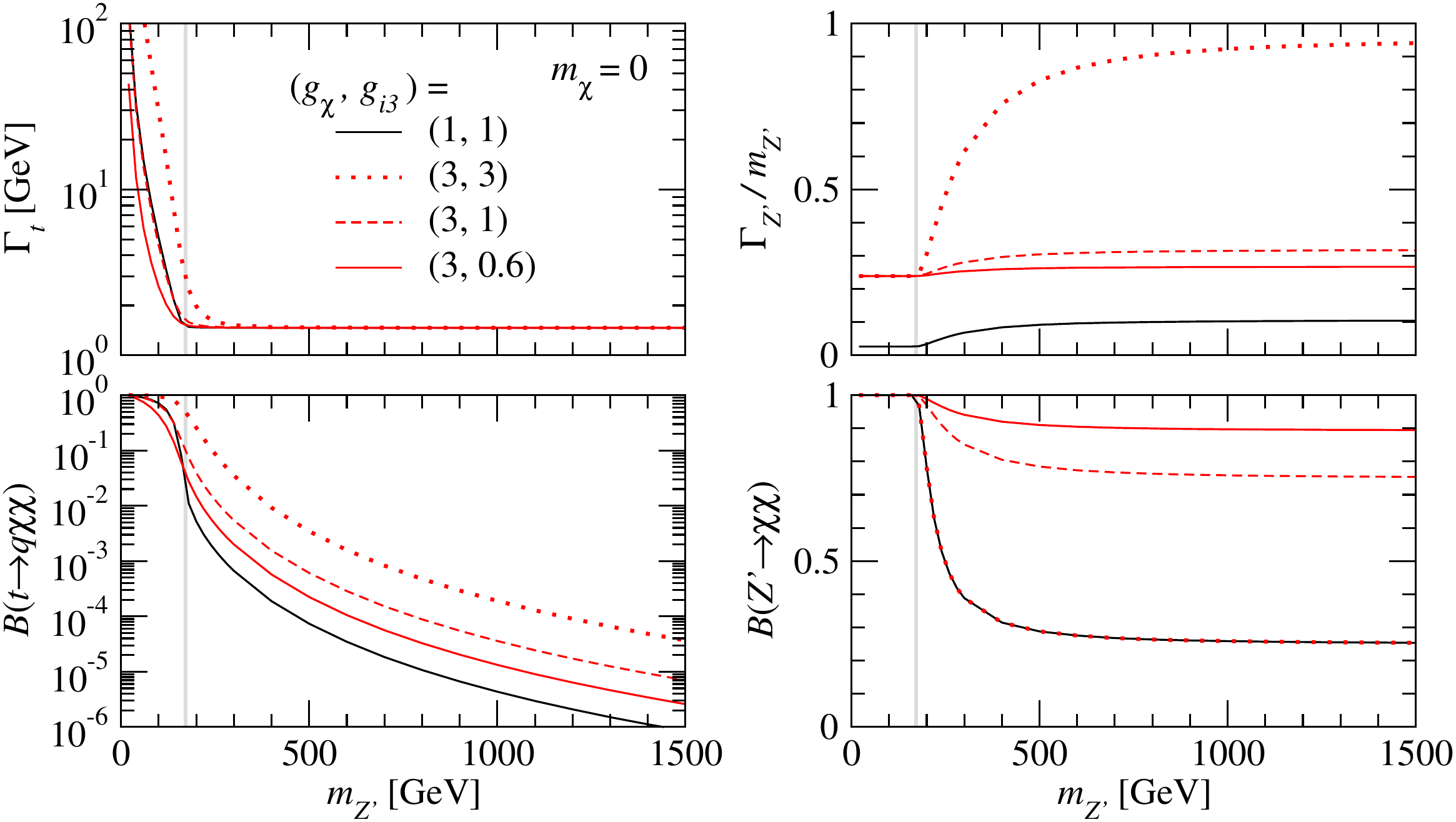}
\caption{\label{fig:br}
 Left: Top decay width and branching ratio of the anomalous top decay as
 a function of the mediator mass.
 Right: $\Gamma_{Z'}/m_{Z'}$ ratio and branching ratio of the mediator
 decay into a pair of DM as a function of the mediator mass.
 We assume a massless DM and take different values of the coupling
 parameters.
 The vertical grey line indicates $m_{Z'}=m_t$.
}    
\end{figure}

\begin{figure}
\center
 \includegraphics[width=0.95\textwidth]{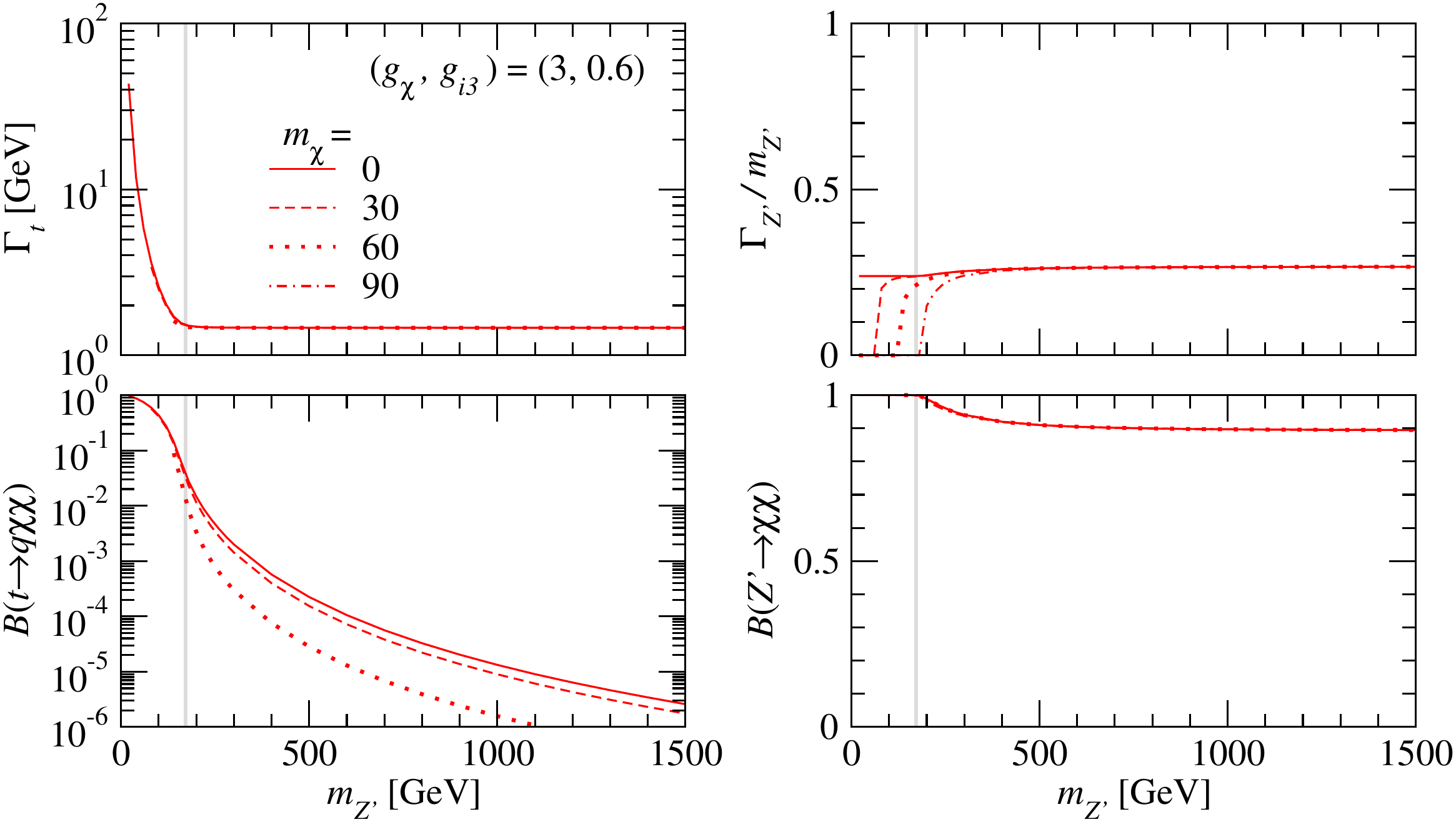}
\caption{\label{fig:br_mx}
 Same as Fig.~\ref{fig:br}, but for the different DM masses.
 The coupling parameters are fixed at $(g_\chi,g_{i3})=(3,0.6)$.
}
\end{figure}

If $m_t>2m_\chi$, top quarks can decay into a light quark (up or charm) 
plus a pair of DM through the on-shell or off-shell mediator:
\begin{align}
 t \to q + Z'^{(*)} \to q + \chi +\bar\chi\,. 
\end{align}
Figure~\ref{fig:br}(left) shows the total width of the top quark (top)
and the branching ratio of the above anomalous decay (bottom) as a
function of the mediator mass, where we assume a massless DM.  
For $m_{Z'}<m_t$, where the top quark decays into the mediator on
mass-shell, the width becomes too broad to be consistent with the
current bound $1\lesssim\Gamma_t\lesssim4$~GeV from
Tevatron~\cite{Abazov:2012vd,Aaltonen:2013kna}. 
We note that in this parameter region the width depends only on the
coupling $g_{i3}$ and the mediator mass.  
For $m_{Z'}>m_t$, on the other hand, the anomalous decay arises through
the off-shell mediator and hence is strongly suppressed as the mediator
mass increases, unless the couplings are very large. 
In Fig.~\ref{fig:br_mx}(left), we show the DM mass dependence with
the coupling fixed at $(g_\chi,g_{i3})=(3,0.6)$.
The anomalous decay becomes smaller when the DM becomes heavier due to
the phase space suppression. 

As a reference, we mention the current limit on the top-quark
flavour-changing neutral current (FCNC) decay, although the analyses
have not considered missing energy.
The current most stringent limit is for the $t\to Zq$ mode as 
$B(t\to Zq)<5\times 10^{-4}$ from the LHC Run-I
data~\cite{Chatrchyan:2013nwa}.

\paragraph{Mediator decay:}

The decay of the mediator depends strongly on the mass spectra and the values of the couplings.  
There are two decay modes, with partial widths given by 
\begin{align}
 \Gamma(Z'\to\chi\bar\chi) &=\frac{g_{\chi}^2 m_{Z'}}{12 \pi} \sqrt{1-\frac{4 m_{\chi}^2}{m_{Z'}^2}} \left(1+2\frac{m_{\chi}^2}{m_{Z'}^2}  \right) \,, \\
 \Gamma(Z'\to t\bar q+\bar tq) &=\frac{ g_{i3}^2 m_{Z'} }{4 \pi} \left( 1-\frac{m_{t}^2}{m_{Z'}^2} \right) \left(  1-\frac{m_t^2}{2 m_{Z'}^2} - \frac{m_t^4}{2 m_{Z'}^4} \right)  \,.
\end{align}

Figures~\ref{fig:br} and \ref{fig:br_mx}(right) show the ratio of the
width to the mass for the mediator (top) and the branching ratio of the 
mediator decay into a pair of DM (bottom) as a function of the mediator
mass.  
For the massless DM (Fig.~\ref{fig:br}), if $m_{Z'}<m_t$, the mediator
can only decay into a pair of DM.
For $m_{Z'}\gg m_t$, on the other hand, the mediator dominantly decays
into a top quark and a light quark if the two couplings are of similar size 
$g_\chi\sim g_{i3}$ due to the colour factor.
For the large $g_{i3}$ coupling the width becomes too large.
Figure~\ref{fig:br_mx}(right) shows that the branching ratio does not 
depend on the DM mass for $m_{Z'}\gg 2m_\chi$.
We note that, even if $Z'$ is in the bottom of the mass spectrum, it
decays via the off-shell top quark (and/or $W$ boson), but its decay is
strongly suppressed.
Moreover, the loop-induced dijet decay channel can be
relevant~\cite{Boucheneb:2014wza}. 

In this work, we are interested in the DM signature at the LHC, and hence we
take $g_\chi=5\times g_{i3}$ so that the branching ratio
$B(Z'\to\chi\bar\chi)$ becomes more than 0.9. 
As our illustrative benchmark point, we take
\begin{align}
\label{couplings_value}
 g_{\chi} =3.0\quad {\rm and}\quad g_{i3} = 0.6\,,
\end{align}
which can provide reasonable signal rate at the LHC as we will show
below, still keeping the mediator width as $\Gamma_{Z'}/m_{Z'}\lesssim 1/4$.

\paragraph{Collider signatures:}

\begin{figure}
\center
 ($qg$)
 \includegraphics[width=0.18\textwidth]{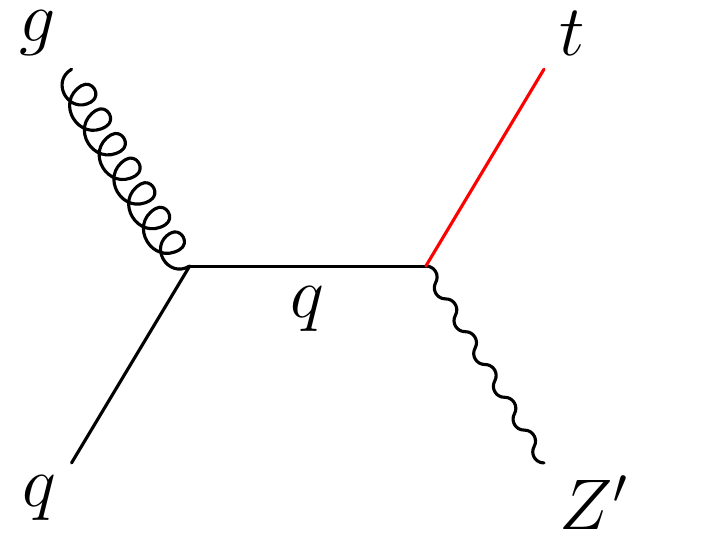}
 \includegraphics[width=0.18\textwidth]{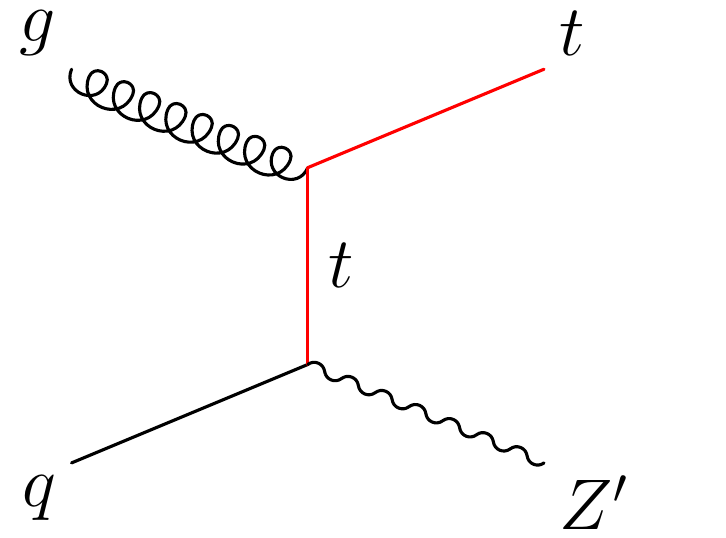}\\[1cm]
 ($qg$)
 \includegraphics[width=0.18\textwidth]{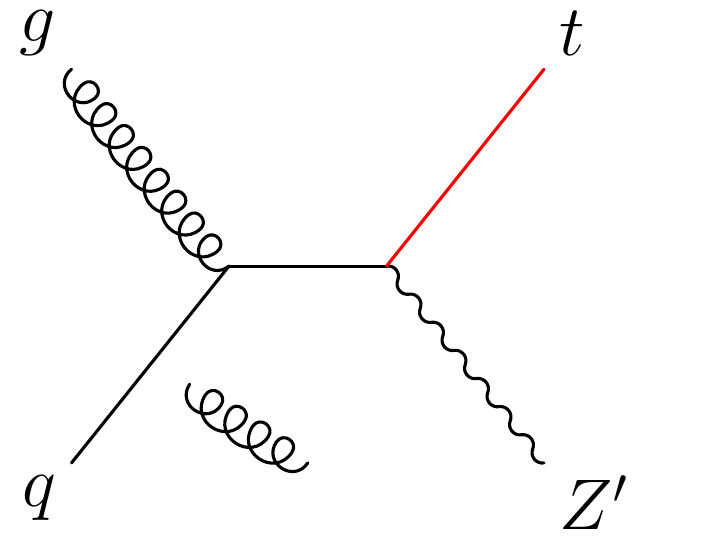}
 \includegraphics[width=0.18\textwidth]{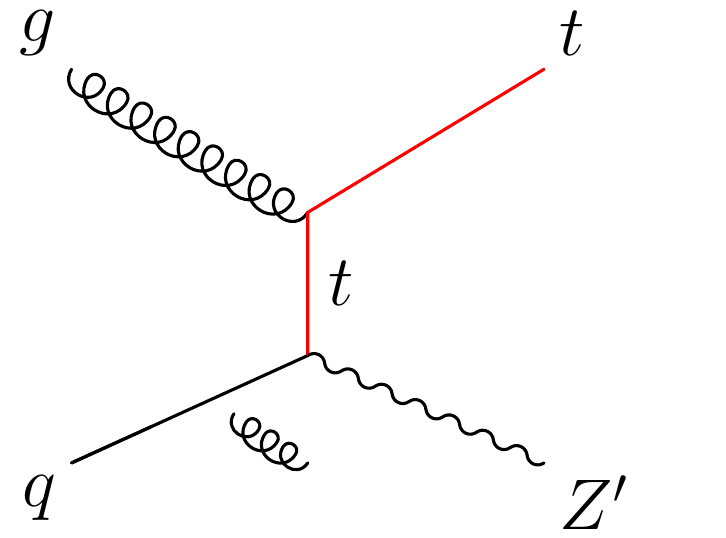}\\
 ($gg$)
 \includegraphics[width=0.18\textwidth]{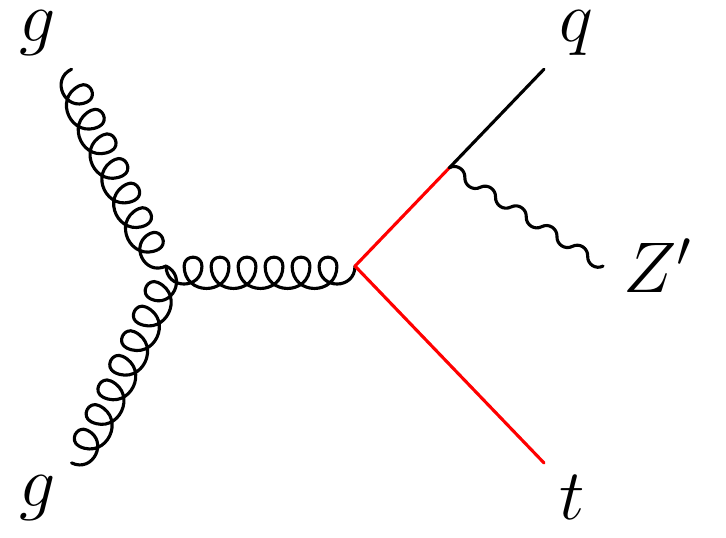}
 \includegraphics[width=0.18\textwidth]{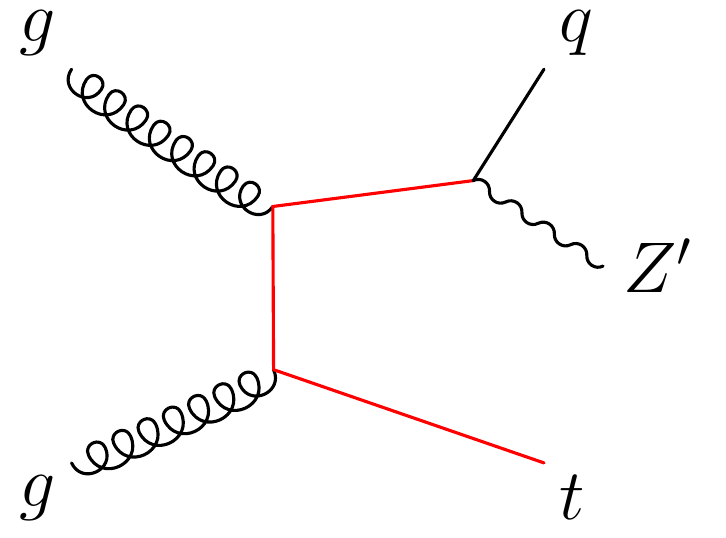}
 \includegraphics[width=0.18\textwidth]{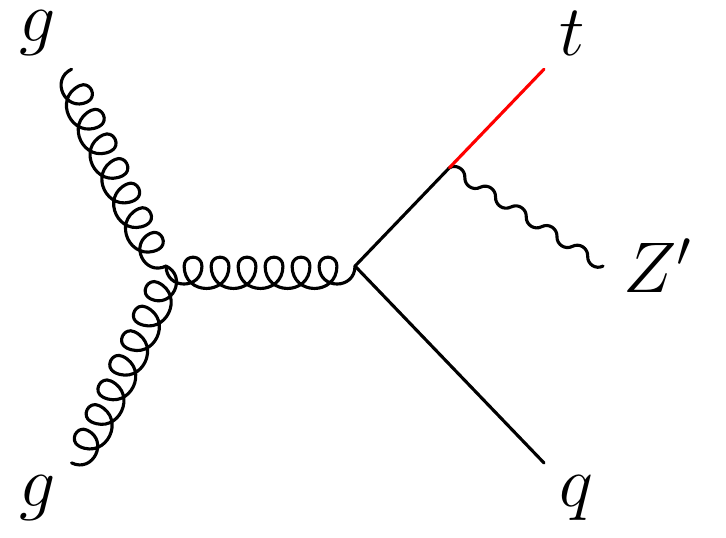}
 \includegraphics[width=0.18\textwidth]{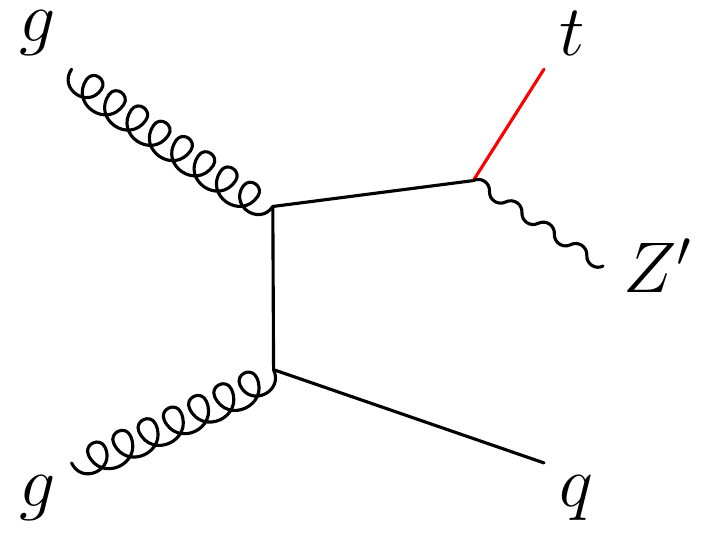}
 \includegraphics[width=0.18\textwidth]{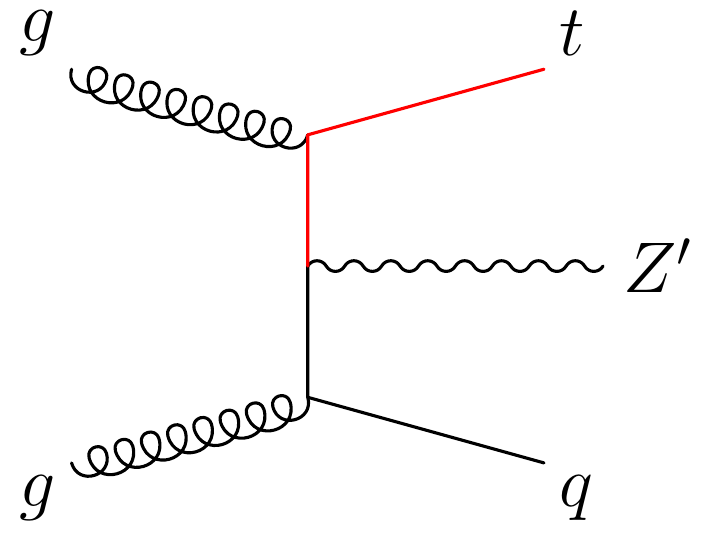}\\
 ($qq$)
 \includegraphics[width=0.18\textwidth]{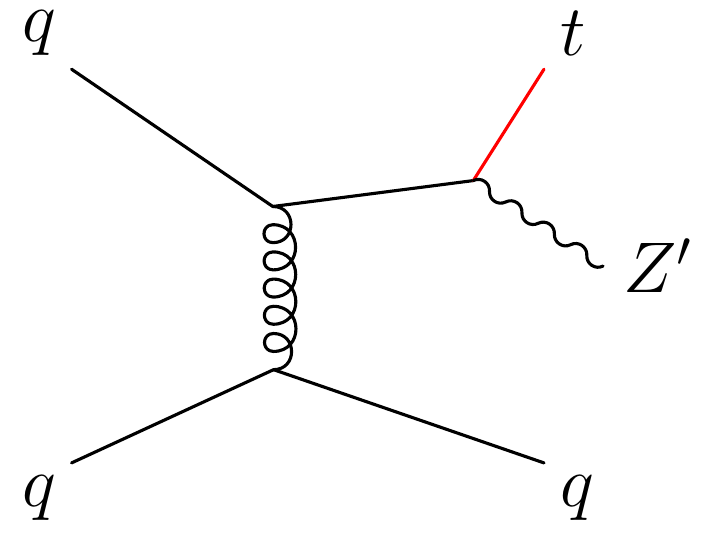}
 \includegraphics[width=0.18\textwidth]{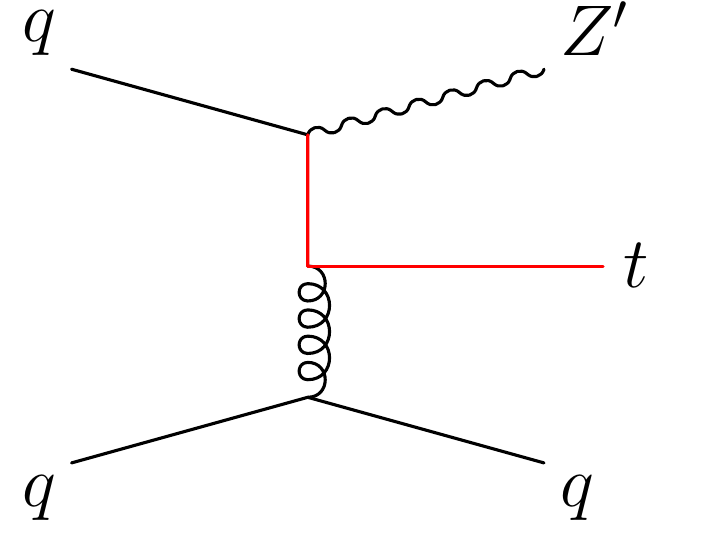}
\caption{\label{fig:diagram}
 Representative Feynman diagrams for $pp\to tZ'$ (top) and $pp\to tZ'j$ (bottom).
}
\end{figure}

A distinctive collider signature in our model is a single top-quark
production in association with large missing energy, the so-called
monotop signature: 
\begin{align} 
 p+p\to t+Z'^{(*)}\to t+\chi+\bar\chi\,,
\label{monotopprocess}
\end{align}
where $t$ denotes a top quark or a top anti-quark. 
The Feynman diagrams are shown in Fig.~\ref{fig:diagram}(top).

\begin{figure}
\center
 \includegraphics[height=.32\textheight]{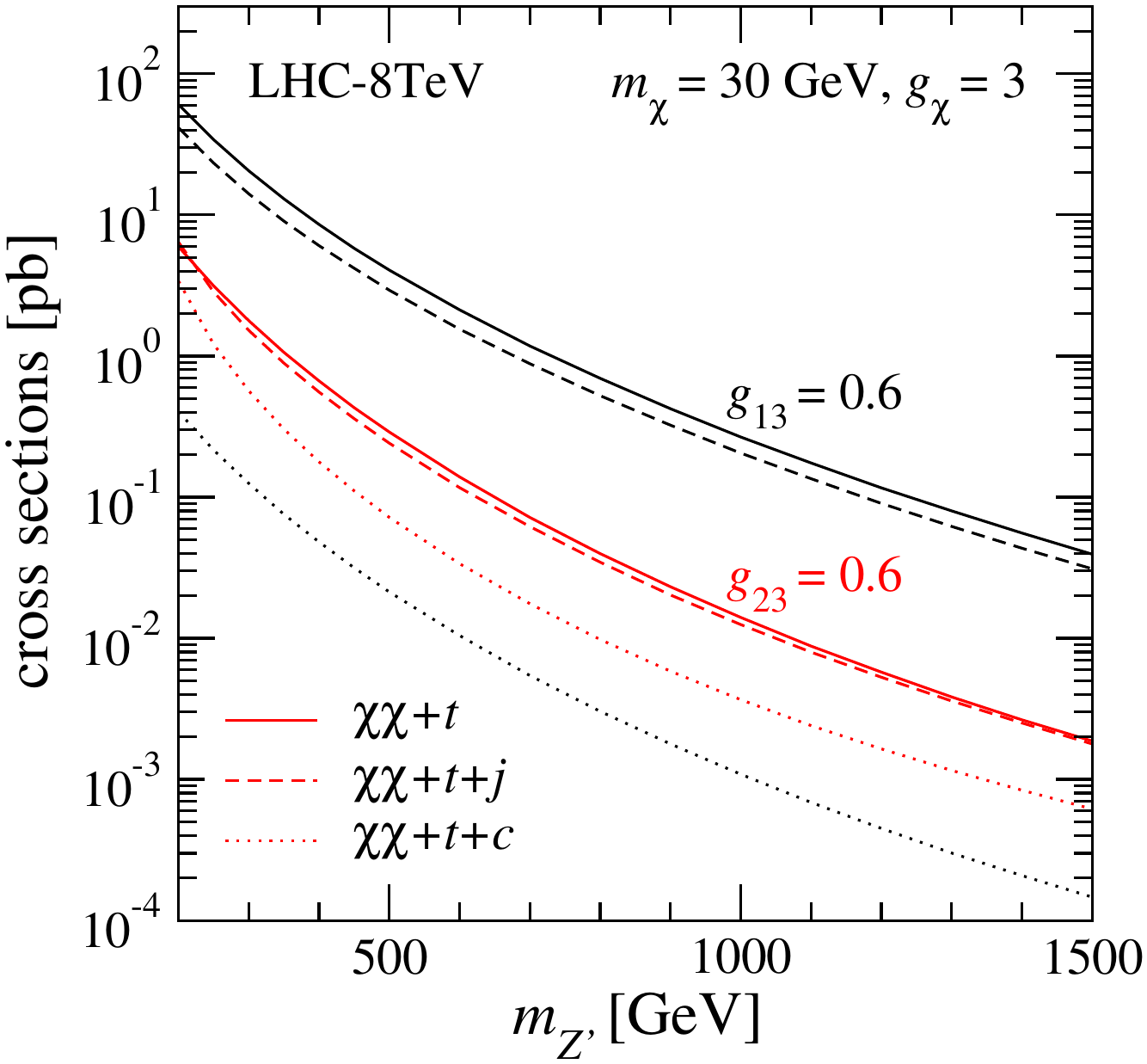}\quad
 \includegraphics[height=.32\textheight]{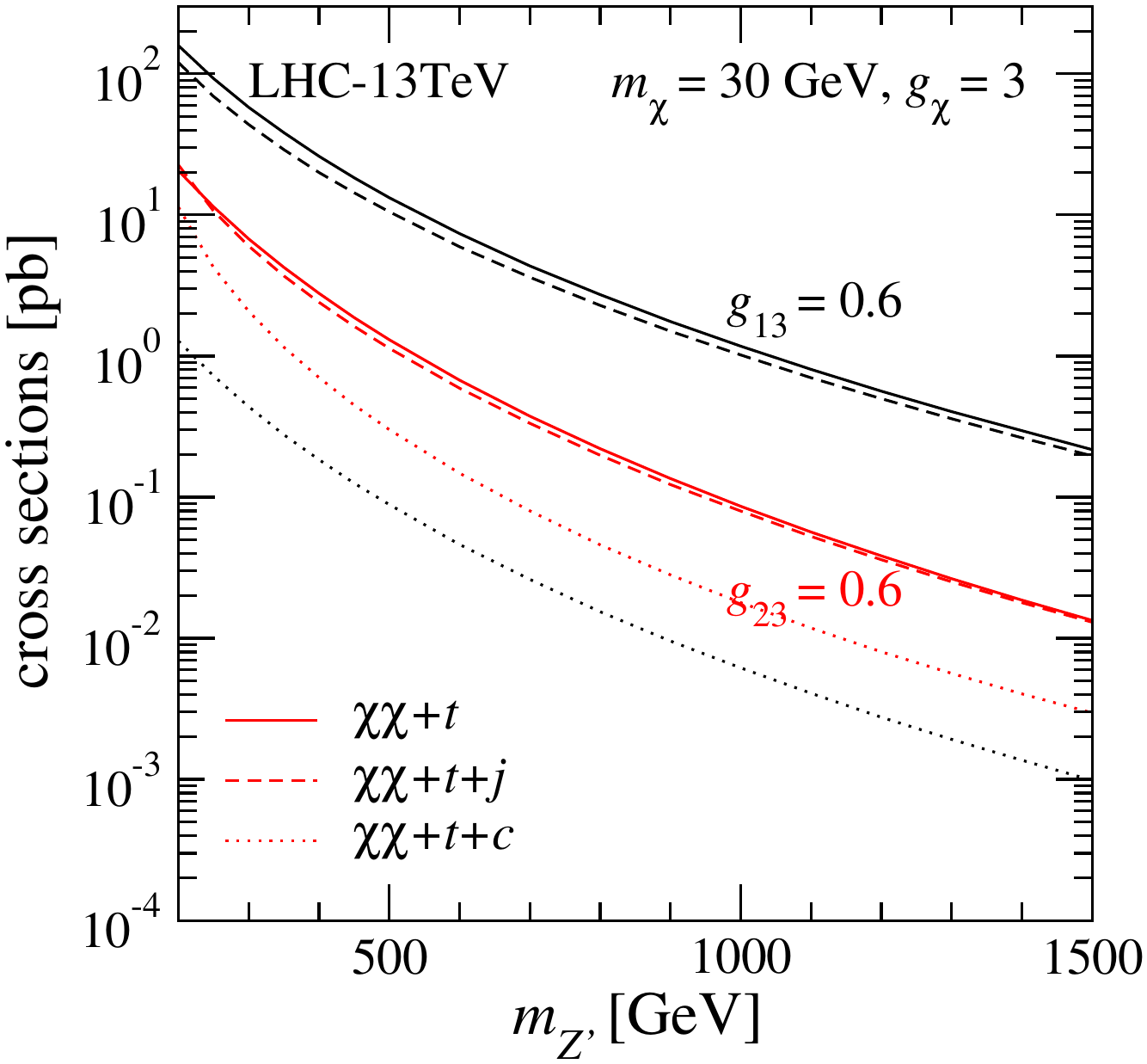}
\caption{\label{fig:xsec}
 Total cross sections for DM pair production in association with a top
 quark in $pp$ collisions at $\sqrt{s}=8$~TeV (left) and 13~TeV (right)
 as a function of the mediator mass, where we assume the top--up (black)
 and top--charm (red) flavour-changing DM model and fix the DM mass at
 30~GeV.  
 For the processes with a ($c$-)jet, the kinematical cuts $p_T^j>25$~GeV 
 and $|\eta^j|<2.5$ are imposed.
}
\end{figure}

Figure~\ref{fig:xsec} shows the total cross sections of 
$pp\to t\chi\bar\chi$ (solid lines) for our benchmark
point~\eqref{couplings_value} at $\sqrt{s}=8$ and 13~TeV as a function
of the mediator mass, where we fix the DM mass at 30~GeV and consider
only $m_{Z'}>m_t$ to avoid the large top width. 
The cross section in the top--up model is larger than that in the
top--charm one roughly by a
factor of 10, simply explained by the difference between the up and
charm parton distribution functions (PDFs).
The cross sections for the both models increase by 3--5 times from
$\sqrt{s}=8$~TeV to 13~TeV. 
We note that the cross sections do not depend on the DM mass as long as 
the mediator is produced on-shell and the $Z'\to\chi\bar\chi$ branching ratio is fixed.

In Fig.~\ref{fig:xsec}, we also show the cross sections for the monotop
process in association with a jet by dashed lines, where we impose the
transverse momentum $p_T^j>25$~GeV and the pseudorapidity
$|\eta^j|<2.5$ as minimal cuts.
As extra QCD jets often emerge at the energy scale of the LHC, we should
take them into account for a reliable prediction, and indeed the cross
sections are comparable with the ones without an additional jet, especially for the
$\sqrt{s}=13$~TeV case. 
As shown in Fig.~\ref{fig:diagram}, in addition to the leading-order
(LO) $qg\to tZ'$ process with a gluon emission, the $gg$ and $qq$
initial states contribute and enhance the production rate.
We note that the steeper fall of the $\chi\bar\chi tj/c$ cross sections
in the top--charm model for around $m_{Z'}\sim 200$~GeV comes from the
top-pair contribution with the anomalous top decay, i.e. 
$\sigma(t\bar t)\times B(t\to q\chi\bar\chi)$; see also
Fig.~\ref{fig:br}(bottom-left).
 
The extra jet contribution can not only enhance the signal but also give
some hint to distinguish between the top--up and top--charm models. 
Although the charm-quark tagging is more difficult than the bottom-quark
tagging, the technique is under development both in the ATLAS and CMS
collaborations and promising for the LHC Run-II.
Assuming the ideal 100\,\% $c$-tagging efficiency, dotted lines in
Fig.~\ref{fig:xsec} present the cross sections for a single top plus a 
charm jet in association with missing transverse energy. 
The production cross section in the top--up model is not zero, but
strongly suppressed, since this comes from the $uc$ initial state only. 
For the top--charm model, on the other hand, the $gg$ scattering can
provide the charm final state, and hence the production rate does not
decrease so much even after identifying the jet as a charm jet. 
In this work, therefore, we take into account extra jets for the
monotop signal by employing a matrix-element parton-shower (ME+PS)
merging scheme~\cite{Alwall:2007fs} and investigate if we can get
additional information on the models.

It should be noted that if $g_{\chi}\lesssim g_{i3}$, i.e. if the 
DM interaction with $Z'$ is subdominant, the dark sector is essentially decoupled and the model becomes a type of non--universal $Z'$ model, such as those intensively discussed in the context of the top forward-backward asymmetry reported by Tevatron~\cite{Aaltonen:2011kc,Abazov:2011rq}.
The $t$-channel $Z'$ produces top quarks in the forward region for
$q\bar q\to t\bar t$~\cite{Jung:2009jz}.
Another distinctive signature in this scenario is same-sign $tt$ pair
production via $qq$ or $\bar q\bar q$ scattering with a $t$-channel
$Z'$~\cite{Berger:2011ua}, searched for already in the LHC Run-I
data~\cite{Chatrchyan:2011dk,Aad:2012bb}.  
The diagrams in Fig.~\ref{fig:diagram} also
produce the same-sign top pair with jets if the $Z'$ dominantly decays
into a top and a light quark.
Note that if the new vector boson is not self-conjugate the model does not
lead to the same-sign top signal \cite{Jung:2011zv}.

%%%%%%%%%%%%%%
\subsection{Top--up vs. top--charm interactions}
%%%%%%%%%%%%%%

As mentioned in Sec.~\ref{intro}, we focus primarily on the less
explored top--charm DM model.
Here, we list certain remarks related to the similarities and differences of the top--up and top--charm flavour-changing DM models, that will be discussed along the paper: 
\begin{itemize}
\item The annihilation of DM (relevant for the calculation of the relic abundance and the indirect detection limits) is practically the same in the top--up and top--charm DM models.

\item DM direct detection physics is a priori different in the two
      models, since the top--up DM model involves the interaction with a
      valence quark in nucleons. 
      However we will demonstrate that both the top--up and top--charm models
      are beyond the reach of current and near future direct detection 
      experiments. 

\item The contribution to the top width and the mediator width is
      once again practically the same in the top--up and top--charm DM models.

\item At the LHC, the main difference between the top--up and top--charm
      DM models lies in the monotop production cross section (if we
      assume $g_{13}\sim g_{23}$); see
      Fig.~\ref{fig:xsec}. 

\item In the monotop signature, as we will explore in
      section~\ref{sec:monotop}, the top--up and top--charm DM models
      can be distinguished by lepton charge asymmetry and by a
      charm-tagging technique on an extra jet.
\end{itemize}

In conclusion, distinguishing between the top--up and top--charm
models is very difficult in non-collider experiments, and may be
challenging at the LHC.
However, in this work we demonstrate that a combined approach can allow us in certain cases to select one of the two models as soon as a hint of new physics is discovered or enough luminosity at the LHC is collected.

\subsection{Effective field theory description}

Before turning to the detailed study of the phenomenology of the DM
model, we introduce the corresponding EFT description to briefly mention
the relation to the simplified model and its validity.

The EFT Lagrangian corresponding to the simplified model
in~\eqref{ZprimeLag} is given by the following four--fermion contact interaction
operators
\begin{align}
 \cL_{int}^{EFT}=\frac{1}{\Lambda^2}\,\ov{\chi}\gamma^{\mu}\chi\,
  ( c_{13}\,\ov{u}_R\gamma_{\mu}t_R 
   +c_{23}\,\ov{c}_R\gamma_{\mu}t_R +h.c.)\,,
\label{eftLag}
\end{align}
where $\Lambda$ is the cutoff scale.

The EFT Lagrangian provides a valid description of the simplified model
in the limit where the mediator is much heavier than the energy scale
probed by the experiment. Therefore, for low-energy processes such as
the DM annihilation in the late universe (relevant for indirect DM
searches) and the elastic scattering of DM off nuclei (relevant for
direct DM searches), the EFT Lagrangian provides an accurate description
of the dynamics. However, if the energy reach is comparable or higher
than the mediator mass such as at the LHC, the EFT approach does not offer a suitable framework for describing the DM interactions~\cite{Goodman:2010yf,Bai:2010hh,Fox:2011pm,Shoemaker:2011vi,Busoni:2013lha,Buchmueller:2013dya}.

\begin{figure}
\center
 \includegraphics[width=.45\textwidth]{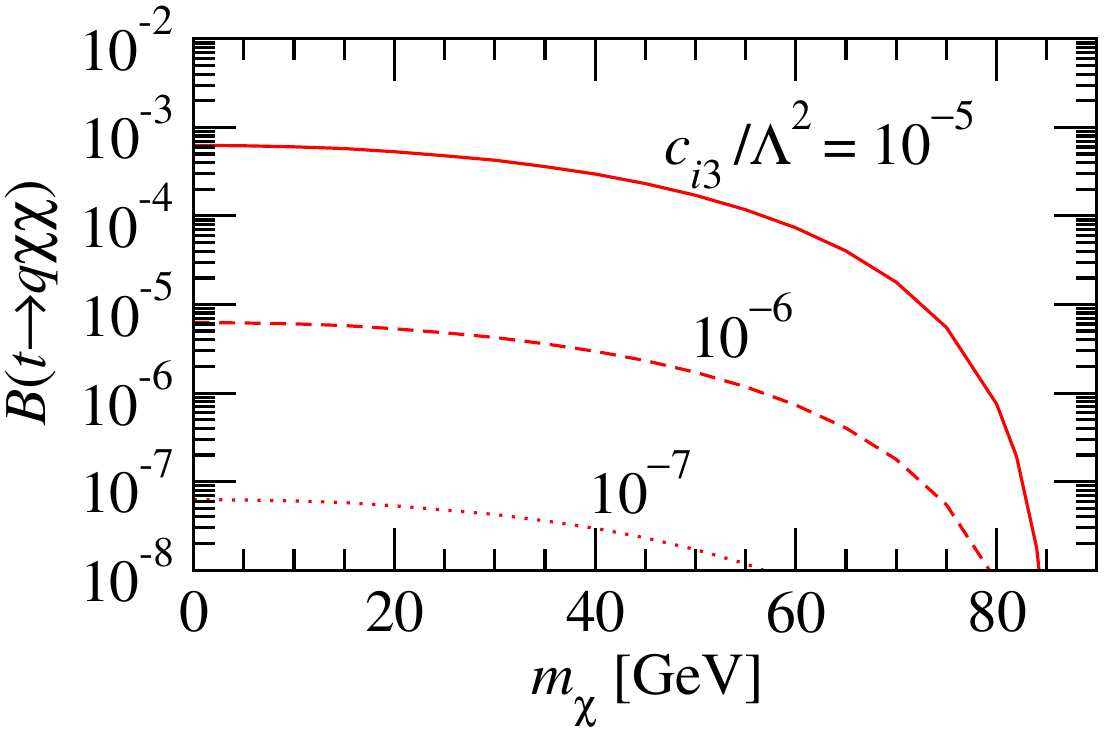}\qquad
 \includegraphics[width=.45\textwidth]{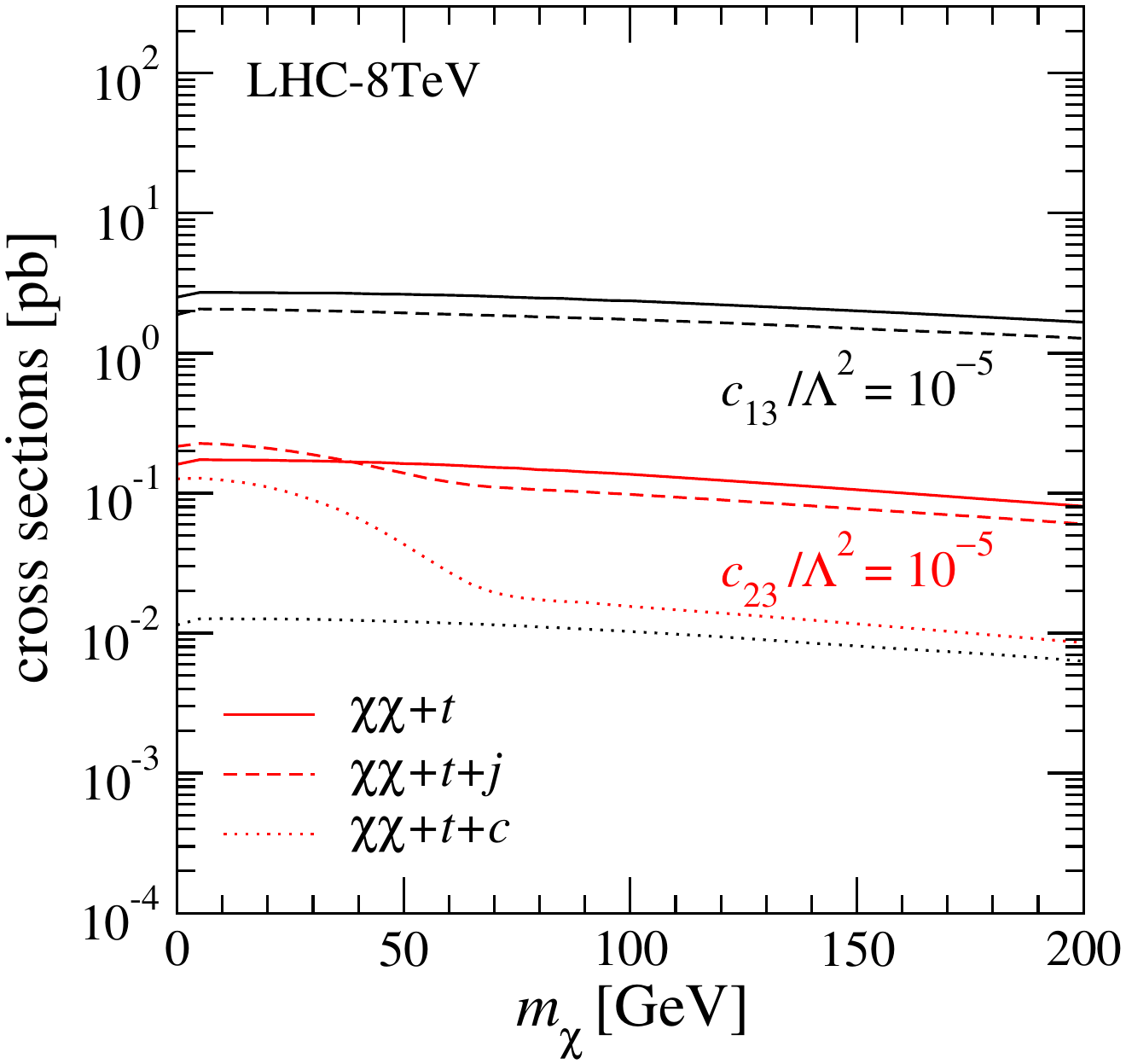}
\caption{\label{fig:eft}
 Left: Branching ratio of the anomalous top decay as a function of the
 DM mass for different coupling parameters in the EFT description.
 Right: Total cross sections for DM pair production in association with
 a top  quark in $pp$ collisions at $\sqrt{s}=8$~TeV as a function of
 the DM mass, where we assume the top--up (black) and top--charm (red)
 flavour-changing DM model in the EFT description.
 For the processes with a ($c$-)jet, the kinematical cuts $p_T^j>25$~GeV 
 and $|\eta^j|<2.5$ are imposed.
} 
\end{figure}

In order to give an idea of the region of the EFT parameter space
that the LHC explores, let us first show the branching ratio
of $t\to q\chi\bar\chi\ (q=u\ {\rm or}\ c)$ in the EFT description
in Fig.~\ref{fig:eft}(left), corresponding to the left-bottom panel in
Figs.~\ref{fig:br} and \ref{fig:br_mx} for the simplified model.
The partial width depends only on $(c_{i3}/\Lambda^2)^2$ and $m_\chi$.
For $m_{\chi} \geq m_t/2$ the decay channel is kinematically closed.
For $c_{i3}/\Lambda^2=10^{-5}$ the anomalous decay branching ratio can
be of the order of $10^{-3}-10^{-4}$. 

Figure~\ref{fig:eft}(right) shows the
production rates for the single top plus missing energy at
$\sqrt{s}=8$~TeV in the EFT description, corresponding
to Fig.~\ref{fig:xsec}(left). $c_{i3}/\Lambda^2$ is fixed at $10^{-5}$. 
The cross sections are insensitive to the DM mass, except for the light
DM case in the top--charm model, where the top-pair production
contributes significantly. 

As we will discuss, the monotop searches at the LHC Run-I set
an upper limit cross section of about
${\cal O}(1)$~pb~\cite{Aad:2014wza,Khachatryan:2014uma}. Hence the EFT
parameter $c_{23}/\Lambda^2\sim {\cal O}(10^{-5})$ is the range that the  
LHC can explore in this model. 
By the tree-level matching relation between the EFT coefficient $c_{i3}$
and the $Z'$ model parameters
\begin{align}
 \frac{c_{i3}}{\Lambda^2}=\frac{g_{\chi}\, g_{i3} }{m_{Z'}^2}\,,
\label{eft_simp}
\end{align}
we can translate the value of the EFT parameter to the simplified model obtaining
\begin{align}
 m_{Z'}\sim
 \begin{cases}
  300~{\rm GeV}
    &\text{with}\quad g_{\chi}\,g_{23}\sim {\cal O}(1)\,,\\
  1~{\rm TeV}
    &\text{with}\quad g_{\chi}\,g_{23}\sim {\cal O}(10)\,, \\
  10~{\rm TeV}
    &\text{with}\quad g_{\chi}\,g_{23}\sim {\cal O}(10^3)\,.
 \end{cases}
\end{align} 
For reasonable values of the couplings, the on-shell production of the mediator is within
the LHC reach, which implies that the EFT is not valid. For the heavy $Z'$ case, on the other hand, the couplings extend beyond the perturbative regime.

In the rest of the paper, we only consider the simplified model for the LHC phenomenology, while we mention the EFT approach in the relic density computation and in indirect DM experiments.

%%%%%%%%%%%%%%%%%%%%
\section{Signatures}\label{sec:signatures}
%%%%%%%%%%%%%%%%%%%%

In this section we study the signatures of the top flavour-changing DM model in detail. First, we discuss the monotop signal for the LHC Run-II. Then, we consider the limits from the non-collider DM experiments. Lastly, we combine the constraints from the collider and non-collider experiments to determine the viable parameter space of the model.

In the following analyses we follow the strategy described in
Ref.~\cite{Christensen:2009jx} for new physics simulations. 
We have implemented the effective Lagrangian~\eqref{ZprimeLag} (as well
as the EFT Lagrangian~\eqref{eftLag}) in
{\sc FeynRules2}~\cite{Alloul:2013bka} to create the model files
interfaced~\cite{Degrande:2011ua,deAquino:2011ub} with
{\sc MadGraph5\_aMC@NLO}~\cite{Alwall:2014hca} for the collider study as
well as with {\sc MicroOMEGAs}~\cite{Belanger:2006is,Belanger:2008sj}
and {\sc MadDM}~\cite{Backovic:2013dpa,Backovic:2015cra} for the
non-collider study.

\subsection{Monotop at the LHC}\label{sec:monotop}

As mentioned in Section~\ref{intro}, the top--up flavour-changing DM
interaction has been studied in the monotop
signature~\cite{Andrea:2011ws,Kamenik:2011nb,Kumar:2013jgb,Wang:2011uxa,Alvarez:2013jqa,Agram:2013wda,Boucheneb:2014wza}
and searched for with the CMS detector for the hadronic top
decays~\cite{Khachatryan:2014uma} and with the ATLAS detector for the leptonic top
decays~\cite{Aad:2014wza}. 

Let us first estimate the constraints on the top--charm flavour-changing DM
interaction from the ATLAS-8TeV analysis~\cite{Aad:2014wza}.
The exotic $v_{\rm met}$ boson in the ``non-resonant'' model in~\cite{Aad:2014wza}
corresponds to the mediator $Z'$ in our model.
They assume that only the top--up coupling is non-zero and the
$v_{\rm met}$ boson decays into invisible particles with 100\,\%
branching ratio.  
They put a bound of about 0.2~pb on the cross section times the
leptonic top-decay branching ratio, 
$\sigma(pp\to tv_{\rm met})\times B(t\to b\ell\nu)$, which is
approximately independent on the $v_{\rm met}$ mass for
$m_{v_{\rm met}}>400$~GeV. 
This translates in an upper bound on the $pp\to tv_{\rm met}$ production cross
section of about 1~pb.
Although we take into account the visible $Z'$ decay, we choose the
coupling in Eq.~\eqref{couplings_value} so that the invisible decay is
dominant, and hence we directly apply the upper limit cross section of
1~pb for the $pp\to t\chi\bar\chi$ cross section in our model.
Figure~\ref{fig:xsec}(left) indicates that the top--up DM model is
bounded to have $m_{Z'}\gtrsim800$~GeV while the top--charm
model is $m_{Z'}\gtrsim400$~GeV for our benchmark couplings.
We note that the cross sections can be rescaled by varying the coupling
$g_{i3}$ and do not depend on the DM mass as long as the mediator is
produced on-shell.

In the following, we will perform a detailed analysis of the LHC Run-II
reach on the monotop signature for the top--charm flavour-changing DM
model.
For the detailed illustration, we take two benchmark points which are
characterized by the light or heavy mediator case: 
\begin{align}
\label{benchmarks}
 A)\ m_{Z'}=400~{\rm GeV}\,, \qquad B)\ m_{Z'}=800~{\rm GeV}\,,
\end{align}
with $m_{\chi}=30$~GeV and $(g_\chi,g_{23})=(3,0.6)$,
and present the kinematical distributions to discuss the selection
cuts. 
We then summarise the signal significance on the $(m_{\chi},m_{Z'})$
plane for a given value of the couplings. 

At the end of this subsection we propose two strategies to distinguish
the top--charm model from the top--up one on the LHC based analyses.
The first one exploits the charge asymmetry of the lepton in the final
state.  
The second one makes use of a charm-tagging technique to distinguish
charm-quark jets from light-quark ($u,d,s$) or gluon jets.

%%%%%%%%%%%%%%%%%
\subsubsection{Prospects for the LHC Run-II}
%%%%%%%%%%%%%%%%%

In this section we study in detail the prospects for discovery of the
top--charm flavour-changing DM model in proton--proton collisions at a
centre-of-mass energy of 13~TeV.

\paragraph{Signal:}
We consider the monotop process~\eqref{monotopprocess} and also take
into account extra jets in the final state. 
In this paper we focus on the leptonic top decay,
$t\to b+W(\to\ell+\nu_{\ell})$, where $\ell$ is an electron or a muon.
Therefore, the signal is characterised by an isolated lepton, a
$b$-tagged jet and extra jets in association with large missing energy.

\paragraph{SM background:}

The following SM backgrounds may mimic the new physics signature:

\begin{itemize}
 \item 
 {\bf Top pair:} 
       The semileptonic decays give rise to the similar final
       state to the signal, and
       this is the main background after the selection
       cuts as shown below. 
       The larger jet multiplicity is expected due to the hadronic decay
       of one of the top quark.
 \item 
 {\bf Single top:}
       Single top production is the only irreducible background.
       Unless it is produced in association with a $W$ boson we expect
       the missing energy to be aligned with the lepton since it
       originates from the same decaying $W$ boson, and hence suitable
       kinematic cuts can reduce this background efficiently. 
 \item 
 {\bf $W$+jets:}
       The production of a $W$ boson (with a leptonic decay) in
       association with jets should also be considered since the total 
       cross section is many orders of magnitude larger than the signal.  
\end{itemize}

The presence of only one lepton in the final state for the signal
removes all processes with $Z$ bosons from the list of relevant
backgrounds.
Furthermore the presence of large missing transverse energy, the jet
multiplicity, and specific angular distributions of the final-state
particles can be exploited to distinguish the signal from the
background.
These specific features are the motivation behind the cuts that will now
be discussed.  

We generate the inclusive signal and SM background samples by employing
the ME+PS merging scheme with {\sc Pythia6}~\cite{Sjostrand:2006za},
implemented in {\sc MadGraph5\_aMC@NLO}~\cite{Alwall:2008qv}.
The fast detector simulation is performed by the {\sc Delphes3}
package~\cite{deFavereau:2013fsa} with the CMS-based detector setup. 
We employ {\sc MadAnalysis5}~\cite{Conte:2012fm,Conte:2014zja} for the
analyses. 
The $t\bar t$ and single-$t$ cross sections are normalised to 831~pb and 
299~pb, respectively~\cite{LHCtopWG:Online}, while the $W$+jets sample
is normalised to $\sigma^{\rm NLO}(W^\pm j)$ of about 
$3\times10^4$~pb~\cite{Alwall:2014hca}.

\paragraph{Event selection}

The final state contains leptons (muons or electrons) and jets as
visible objects. 
Leptons are required to be isolated.%
\footnote{All the energy surrounding the lepton in the cone 
($\Delta R=0.4$) divided by the lepton $p_T$ is below 0.2.} 
Jets are reconstructed by employing the anti-$k_T$
algorithm~\cite{Cacciari:2008gp} with a radius parameter of 0.5. 
Leptons and jets are required to have $p_T>30$~GeV and $|\eta|<2.4$.

\begin{figure}
\center
 \includegraphics[width=.48\textwidth]{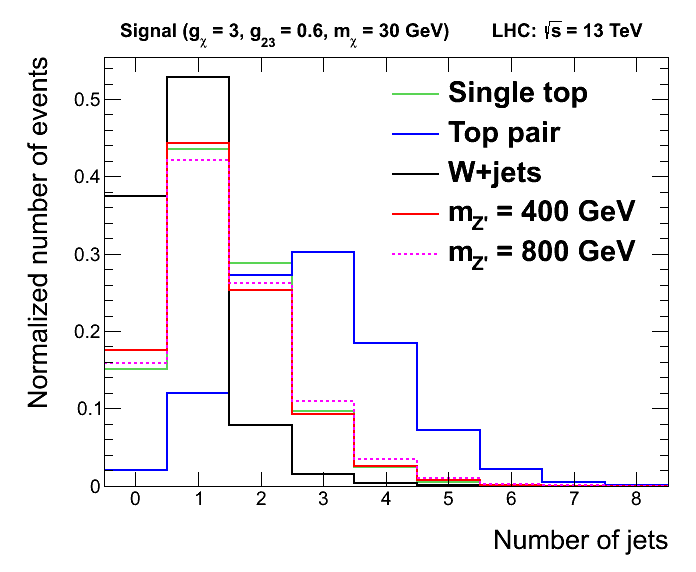}\quad
 \includegraphics[width=.48\textwidth]{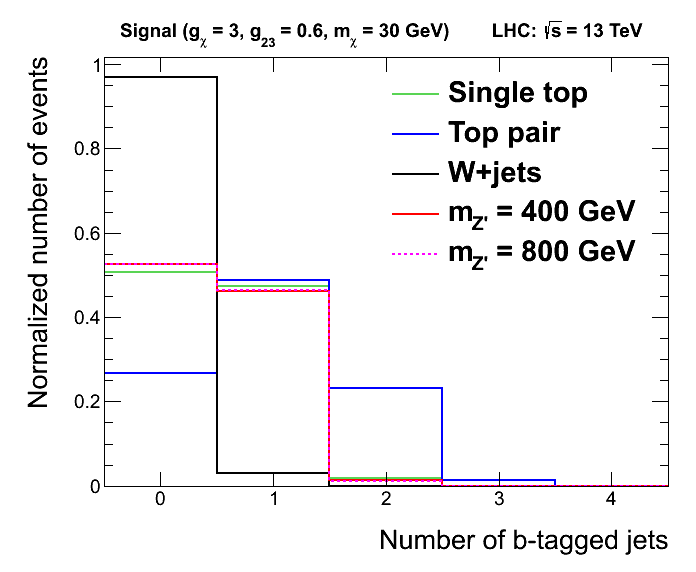}\\[2mm]
 \includegraphics[width=.48\textwidth]{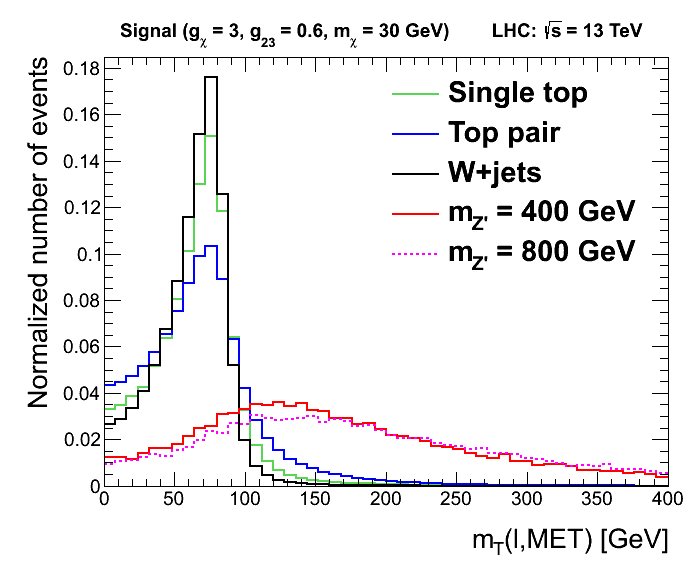}\quad
 \includegraphics[width=.48\textwidth]{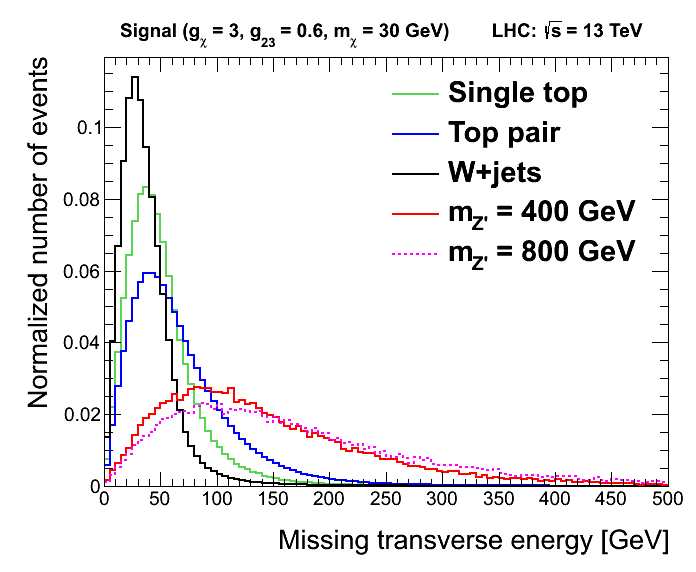}\\[2mm]
 \includegraphics[width=.48\textwidth]{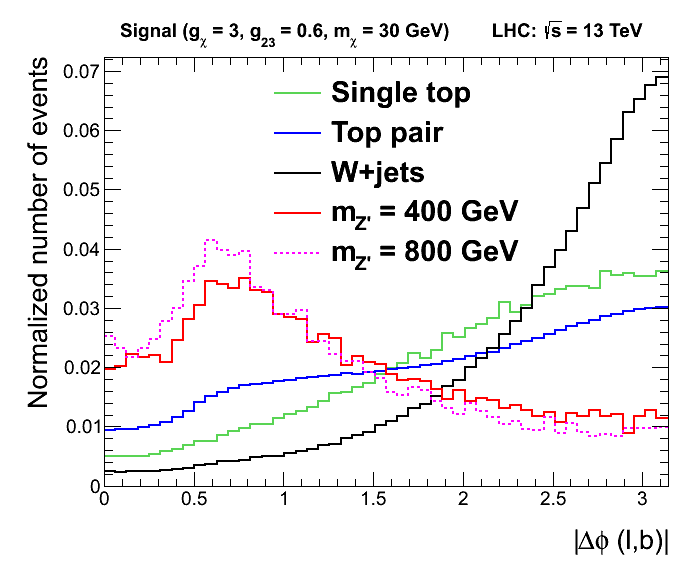}
\caption{\label{fig:distributions}
 Normalised distributions of the variables used in the analysis for the
 signal (benchmark point $A$ and $B$) and the SM backgrounds after the
 pre-selection, i.e. we require only $N_\ell=1$.
 For the $\Delta\phi(\ell,b)$ distribution $N_b=1$ is also required.
}
\end{figure}

We pre-select the events by demanding exactly one isolated lepton,
$N_{\ell}=1$.
In Fig.~\ref{fig:distributions} we show distributions for the number of
jets and some kinematical variables,% 
\footnote{The transverse mass is defined as 
$m_{T}(\ell,\MET)=\sqrt{2p_{T}^{\ell}\MET(1-\cos\Delta\phi(\ell,\MET))}$.}
which the ATLAS analysis uses, both
for the SM background and the signal (benchmarks $A$ and $B$) after
having required $N_{\ell}=1$.   

The following set of cuts are employed in order to maximise the
significance of the signal: 
\begin{align}
&N_{\ell}=1 \,,\quad
 1\leq N_{j} \leq 2\,, \quad
 N_{b}=1\,, \nonumber \\
&m_T(\ell,\MET) > m_T^{\text{min}} \,, \quad
 \MET > \MET^{\text{min}} \,, \quad
 |\Delta \phi(\ell,b)| < \Delta \phi^{\text{max}}\,.
\end{align}
\begin{itemize}
\item $1\leq N_{j}\leq 2$:
      The ``pure'' monotop signal contains exactly one jet which should
      be $b$-tagged, coming from the top decay, and the most of the
      previous works including the ATLAS analysis select only the
      one-jet sample. 
      Here, as discussed, we propose to include an extra jet to enhance
      the signal and to utilise it to distinguish the top--charm model
      from the top--up one. 
      The selection is still efficient to reduce the $t\bar t$
      background in which the average jet multiplicity is higher; see
      Fig.~\ref{fig:distributions}(top-left). 
      For convenience, we define two signal regions (SRs):
\begin{align}
 {\rm SR1})\ N_j=1\ {\rm and}\ N_b=1\,,\qquad
 {\rm SR2})\ N_j=2\ {\rm and}\ N_b=1\,.
\label{SR}
\end{align}

\item $N_b=1$: The signal contains one $b$-tagged jet.
      Requiring exactly one $b$-tagged jet reduces the $W$+jets
      background since processes with a $W$ boson in association with
      bottom quarks are rare, and reduces also the $t\ov{t}$ background
      where there is a second jet expected to originate from a bottom
      quark; see Fig.~\ref{fig:distributions}(top-right). 
      For $b$-tagging we use a parametrisation of the efficiency of 
      the combined secondary vertex (CSV) algorithm of the CMS collaboration~\cite{CMS-BTV-12-001}, 
      as a function of the $p_T$, $\eta$ and flavour of the jet. 
      We employ the so-called medium operating point, 
      which overall results in $b$-tagging efficiencies of about 70\,\% for $b$-flavour jets, about 1\,\% for $u$,$d$,$s$-flavour and gluon jets, and about 20\,\% for $c$-flavour jets.

\item  $m_T(\ell,\MET)>m_T^{\text{min}}$: The $m_T(\ell,\MET)$
       distributions in Fig.~\ref{fig:distributions} display a
       remarkable shape difference between the DM signal and the SM
       background.  
       For the SM background the $\MET$ and the lepton originates from
       the same $W$ boson and as a consequence the $m_T(\ell,\MET)$
       distribution drops around the $W$ boson mass.
       This is not the case for the DM signal, where the $\MET$
       originates from the invisible $Z'$ decay.
       The heavier $Z'$ case (benchmark $B$) presents slightly larger
       $m_T$ distribution.

\item $\MET>\MET^{\text{min}}$: The presence of DM in the signal
      introduces a lot of missing energy in the detector. The missing
      transverse energy will therefore be much larger on average for the
      signal in comparison to the SM background processes in which the
      $\MET$ originates from neutrinos only.  

\item $|\Delta \phi(\ell,b)|<\Delta\phi^{\text{max}}$: In the DM signal,
      the lepton and the $b$-jet always originate from the decay of one
      top quark and hence they display a small azimuthal angle
      separation.
      Instead, the SM backgrounds can also present events where the
      $b$-jet and the lepton arise from different decay chains.
      In particular, in the $W$+jets background, the lepton and the
      $b$-jet are most likely back to back.  
\end{itemize}

\paragraph{Results and discovery reach:}

We now present the results of our analysis and the discovery
reach of the LHC-13TeV with integrated luminosity
$\mathcal{L}=100~\text{fb}^{-1}$. 

Table~\ref{tab:cutflow} shows the visible cross sections for the SM
backgrounds and the signal for benchmark point $A$ and $B$ for each
consecutive cut up to the $N_b=1$ selection.
Although the $t\bar t$ and single $t$ backgrounds reduce by a factor of
ten and the $W$+jets drops by a factor of a hundred, the background is still
larger than the signal.

\begin{table}
\center
\begin{tabular}{|l||l|l|l|l|l|}
\hline
 Cuts & Top pair & Single top & $W$+jets & $A$: $m_{Z'400}$ & $B$: $m_{Z'800}$  \\
\hline\hline
$N_\ell=1$ & 1.56 $\times 10^{2}$ & 3.33 $\times 10^{1}$ & 3.00 $\times 10^{3}$
        & $3.74\times10^{-1}$ & $2.96\times10^{-2}$ \\
\hline
$N_j\geq 1$ & 1.53 $\times 10^{2}$ & 2.83 $\times 10^{1}$ & 1.98 $\times 10^{3}$
        & $3.08\times10^{-1}$ & $2.49\times10^{-2}$ \\
\hline
$N_j\leq 2$ & 6.01 $\times 10^{1}$ & 2.41 $\times 10^{1}$ & 1.93 $\times 10^{3}$ 
        & $2.60\times10^{-1}$ & $2.02\times10^{-2}$ \\
\hline
$N_b=1$ & 3.15 $\times 10^{1}$ & 1.33 $\times 10^{1}$ & 8.71 $\times 10^{1}$ 
        & $1.45\times10^{-1}$ & $1.10\times10^{-2}$ \\
\hline
\end{tabular}
\caption{\label{tab:cutflow}
 Cutflow table showing the visible cross sections in pb for the SM
 backgrounds and the signal for benchmark point $A$ and $B$,
 consecutively applying the cuts outlined in the text.
}
\end{table}

We now use the information of the kinematical distributions shown in
Fig.~\ref{fig:distributions}, and maximise the statistical signal
significance 
\begin{align}
\label{stat_sign}
 \mathcal{S} \equiv \frac{S}{\sqrt{S+B}}
\end{align}
to find each optimal cut in the rest of the selection steps.
We find that
\begin{align}
 m_T^{\text{min}} = 150~\text{GeV}\,,\quad  
 \MET^{\text{min}} = 200~\text{GeV}\,,\quad  
 \Delta\phi^{\text{max}} = 1.6\,.
\end{align}
Table~\ref{tab:S1} presents the continuation of Table~\ref{tab:cutflow}
together with the signal significance $\mathcal{S}$ assuming the
integrated luminosity $\mathcal{L}=100$~fb$^{-1}$.
The sensitivities of each of the two signal regions~\eqref{SR} are also shown at
the bottom of the table. 
The main SM background after all the cuts is $t\bar t$.
We can easily obtain the significance larger than $5$ for the light
mediator case (benchmark $A$) with 100~fb$^{-1}$. 
The heavier case (benchmark $B$) is instead at reach to be excluded. 
It is important to note that the signal significance is larger than the
pure monotop sample (SR1) when we include an extra jet in the analysis,
i.e. SR2.    
We also note that the shape of the distributions slightly depends on the 
value of the $Z'$ mass. 
In particular the $m_T$ and $\MET$ distributions for heavier $Z'$ are
centred around larger values.  
Hence the efficiency of these cuts for the benchmark point $B$ is
slightly higher. 
 
\begin{table}
\begin{footnotesize}
\begin{tabular}{|l||l|l|l|lr|lr|}
\hline
 Cuts & Top pair & Single top & $W$+jets 
 & $A$: $m_{Z'400}$ & ($\mathcal{S}$)
 & $B$: $m_{Z'800}$ & ($\mathcal{S}$) \\
\hline\hline
$N_b=1$ & 3.15 $\times 10^{1}$ & 1.33 $\times 10^{1}$ & 8.71 $\times 10^{1}$  & $1.45\times10^{-1}$ & (3.98) 
        & $1.10\times10^{-2}$ & (0.30) \\
\hline
$m_{T}\!>\!150$~GeV & 1.86 $\times 10^{0}$ & 1.74 $\times 10^{-1}$ & 3.25 $\times 10^{-1}$ & $7.83\times10^{-2}$ & (15.86) 
        & $6.76\times10^{-3}$ & (1.39) \\
\hline
$\MET\!>\!200$~GeV & 1.19 $\times 10^{-1}$ & 6.48 $\times 10^{-3}$ & 5.17 $\times 10^{-3}$ & $3.21\times10^{-2}$ & (25.18)
        & $3.38\times10^{-3}$ & (2.92) \\
\hline
$|\Delta\phi|<1.6$ & 8.61 $\times 10^{-2}$ & 4.97 $\times 10^{-3}$ & 2.27 $\times 10^{-3}$  & $3.08\times10^{-2}$ & (27.62)
        & $3.24\times10^{-3}$ & (3.30) \\
\hline\hline
SR1: $N_j=1$ & 3.76 $\times 10^{-2}$ & 2.30 $\times 10^{-3}$ & 1.29 $\times 10^{-3}$ & $1.77\times10^{-2}$& (23.10) 
        & $1.82\times10^{-3}$& (2.77) \\
\hline
SR2: $N_j=2$ & 4.86 $\times 10^{-2}$ & 2.67 $\times 10^{-3}$ & 9.82 $\times 10^{-4}$  & $1.30\times 10^{-2}$& (16.15) 
        & $1.43\times10^{-3}$& (1.95) \\
\hline
\end{tabular}
\caption{\label{tab:S1}
Cutflow table showing the visible cross sections in pb for the SM
 backgrounds and the signal for benchmark point $A$ and $B$, after the
 $N_b=1$ selection.
 The columns $\mathcal{S}$ show the statistical signal
 significance~\eqref{stat_sign} for $\mathcal{L}=100$~fb$^{-1}$.
}
\end{footnotesize}
\end{table}

\begin{table}
\begin{footnotesize}
\begin{tabular}{|l||l|l|l|lr|lr|}
\hline
 Cuts & Top pair & Single top & $W$+jets 
 & $A$: $m_{Z'400}$ & ($\mathcal{S'}$) 
 & $B$: $m_{Z'800}$ & ($\mathcal{S'}$) \\
\hline\hline
$N_b=1$ & 3.15 $\times 10^{1}$ & 1.33 $\times 10^{1}$ & 8.71 $\times 10^{1}$ & $1.45\times10^{-1}$ & (0.01) 
        & $1.10\times10^{-2}$ & (0.00) \\
\hline
$m_{T}\!>\!300$~GeV & 1.09 $\times 10^{-1}$ & 1.24 $\times 10^{-2}$ & 1.19 $\times 10^{-2}$ & 2.15 $\times 10^{-2}$ & (1.61) 
        & 2.43 $\times 10^{-3}$ & (0.18) \\
\hline
$\MET\!>\!350$~GeV & 3.38 $\times 10^{-3}$ & 1.74 $\times 10^{-4}$ & 5.84 $\times 10^{-5}$ & 4.04 $\times 10^{-3}$ & (8.89)
        & 6.69 $\times 10^{-4}$ & (1.61) \\
\hline
$|\Delta\phi|<1.0$ & 1.87 $\times 10^{-3}$ & 1.28 $\times 10^{-4}$ & 0.00 $\times 10^{0}$ & 3.76 $\times 10^{-3}$ & (12.03)
        & 6.07 $\times 10^{-4}$ & (2.36) \\
\hline\hline
SR1: $N_j=1$ & 6.42 $\times 10^{-4}$ & 6.41 $\times 10^{-5}$ & 0.00 $\times 10^{0}$ & 1.75 $\times 10^{-3}$ & (10.09) 
        & 2.76 $\times 10^{-4}$ & (2.27) \\
\hline
SR2: $N_j=2$ & 1.23 $\times 10^{-3}$ & 6.41 $\times 10^{-5}$ & 0.00 $\times 10^{0}$ & 2.00 $\times 10^{-3}$ & (8.99) 
        & 3.31 $\times 10^{-4}$ & (1.82) \\
\hline
\end{tabular}
\caption{\label{tab:S2}
 Same as Table~\ref{tab:S1}, but with the tighter cuts. 
 The columns labelled $\mathcal{S}'$ show the signal significance
 including a 10\,\% systematic uncertainty~\eqref{syst_sign} for
 $\mathcal{L}=100$~fb$^{-1}$. 
}
\end{footnotesize}
\end{table}

The statistical significance $\mathcal{S}$ is not the only
representative for an analysis which has to cope with systematic
uncertainties, such as the ones on the cross sections of the SM
background processes. 
For instance, given that the $t\bar{t}$ is the most relevant SM
background, a large systematic uncertainty will originate from the
uncertainty on the $t\bar{t}$ cross section, which was estimated of
about 7\,\% at $\sqrt{s}=8$~TeV~\cite{Chatrchyan:2013faa}.
Considering also other sources of systematic uncertainties, we
conservatively assume a 10\,\% uncertainty in the SM background 
estimation and define the significance as
\begin{align}
\label{syst_sign}
 \mathcal{S'}=\frac{S}{\sqrt{S+B+(0.1B)^{2}}}\,.
\end{align}

We repeat the same procedure, but maximise the
significance~\eqref{syst_sign} to find a new set of optimal cuts. 
We find a tighter set of cuts
\begin{align}
 m_T^{\text{min}} = 300~\text{GeV}\,,\quad  
 \MET^{\text{min}} = 350~\text{GeV}\,,\quad  
 \Delta\phi^{\text{max}} = 1.0\,.
\end{align}
Table~\ref{tab:S2} gives the results of these selection cuts, and
clearly shows that even after including systematic uncertainties the
potential remains to discover the top--charm flavour-changing DM
events during the run of the LHC with 13~TeV proton--proton collisions
and an expected 100~fb$^{-1}$ of collected data. 

\begin{figure}
\center
 \includegraphics[width=0.5\textwidth]{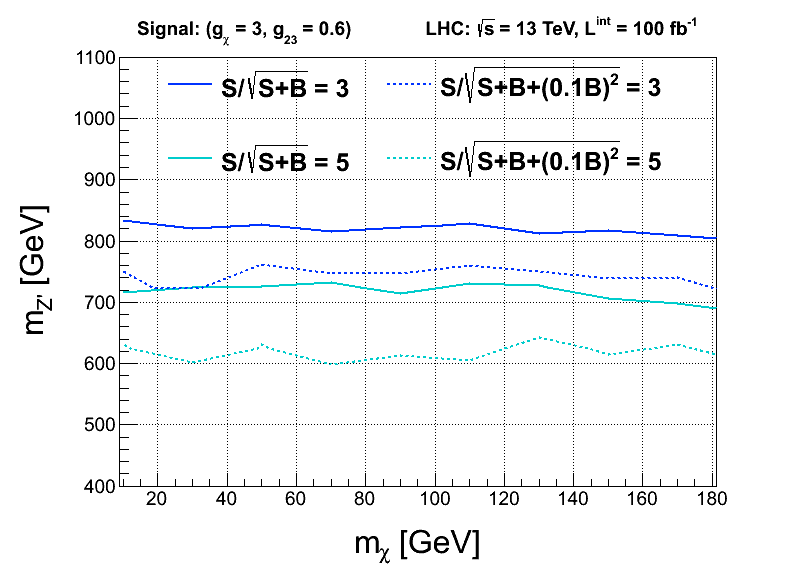}
\caption{\label{fig:ScanPlot}
 3 and 5 $\sigma$ contours of the signal significance in the
 $m_{\chi}-m_{Z'}$ parameter space without (solid) and with (dotted)
 including a systematic uncertainty in the calculation of the
 significance. 
}
\end{figure}

In order to establish the reach of the LHC in the parameter space of the
model, we perform a parameter scan in the $m_{\chi}-m_{Z'}$ plane. 
Figure~\ref{fig:ScanPlot} shows the three sigma and five sigma contours
of the signal significance without ($\mathcal{S}$) and with
($\mathcal{S'}$) a systematic uncertainty in this mass plane. 
There is essentially no dependence on the DM mass since the mediator is
always produced on-shell in this mass range, and
subsequently decaying into a pair of the DM particles. 
On the other hand, the LHC reach is largely dependent on the mediator
mass that determines the production cross section for a given coupling. 
We find that a large part of the mass space is accessible in the 13~TeV
run of the LHC for the reasonable choice of the coupling parameter
such as $(g_\chi,g_{23})=(3,0.6)$.

Following the same analysis strategy, one can easily explore the
corresponding top--up flavour-changing DM model, setting $g_{\chi}=3.0$
and $g_{13}=0.6$ and varying the DM and the $Z'$ masses.
The enhancement of the production cross section due to the up-quark PDF 
(see Fig.~\ref{fig:xsec}) determines a much higher reach for the top--up
model with respect to the top--charm model for analogous values of the
couplings.
Indeed we find that the top--up DM model can be discovered at the
LHC-13TeV for a $Z'$ mass up to about 1.5~TeV.

\subsubsection{Top--charm vs. top--up in monotop}

As discussed, the top--charm and top--up flavour-changing DM models
essentially give the same monotop signature at the LHC. 
The main difference is the overall cross sections, and hence the mass
reach is different if we assume the same couplings between the two
models. 
However, there is no other direct observable which is related to the
$Z'$ mass, since we have seen in Fig.~\ref{fig:distributions} that the
kinematical distributions are similar between the different $Z'$
mass. 
Hence, even if the monotop signal is discovered at the LHC Run-II, it
may be very difficult to discriminate between the top--charm and the
top--up DM models. 

In this subsection we propose possible techniques to distinguish between
the two models in the monotop signature, not based on the overall signal
cross sections.
For this purpose we define two benchmarks for the
top--up DM model:
\begin{align}
 &A)\quad \{g_{\chi},\,g_{13},\,m_{\chi},\,m_{Z'}\}
     =\{3.0,\,0.19,\,30~{\rm GeV},\,400~{\rm GeV}\}\,, \nn\\
 &B)\quad \{g_{\chi},\,g_{13},\,m_{\chi},\,m_{Z'}\}
     =\{3.0,\,0.19,\,30~{\rm GeV},\,800~{\rm GeV}\}\,,
\label{benchmarks_up}
\end{align}
where we choose the same parameters as in the benchmarks of the
top--charm model~\eqref{benchmarks} except the $g_{13}$ coupling.
The value of the coupling $g_{13}$ is chosen so that the top--up monotop
cross sections become comparable to the top--charm ones.

\paragraph{Lepton charge asymmetry:}

The first strategy that we adopt to distinguish the top--charm
flavour-changing DM model from the top--up one is to exploit the lepton
charge asymmetry in the leptonic monotop final state.
Since the up-quark PDF is much larger than the up-antiquark one in
protons, the monotop process for the top--up DM model in proton--proton
collisions produces much more top quarks than top anti-quarks, leading
to a large majority of events with a positively charged
lepton~\cite{Kumar:2013jgb}.  
For the top--charm model, on the other hand, we expect an equivalent
number of events with a positively and negatively charged lepton as the
charm PDF is equal to the charm-antiquark one.

In order to quantify this observation, we look at the signal
significance again, but for positively and negatively charged
leptons separately.  
Note that the main SM background, i.e. $t\bar t$, is charge symmetric.
In Table~\ref{tab:lepton_charge} we report the result of these
investigations for both the top--charm~\eqref{benchmarks} and the 
top--up~\eqref{benchmarks_up} benchmarks. 
We display the significance (both without and with a systematic
uncertainty) for the combined signal region SR1+SR2 of each benchmark
depending on the lepton charge selection. 
The $\ell^+ + \ell^-$ columns correspond to the analysis discussed in the
previous section, which does not distinguish the lepton charge. 
 
\begin{table}
\center
\begin{tabular}{|l|r|r|r||r|r|r|}
\cline{2-7}
   \multicolumn{1}{c}{}  & \multicolumn{3}{|c||}{Top--charm model} 
 & \multicolumn{3}{c|}{Top--up model} \\
\cline{2-7}
  \multicolumn{1}{c|}{} & $\ell^{+}$ + $\ell^{-}$ & $\ell^{+}$ & $\ell^{-}$ & $\ell^{+}$ + $\ell^{-}$ & $\ell^{+}$ & $\ell^{-}$  \\
\hline
$\mathcal{S}$ ($A$: $m_{Z'}=400$~GeV) & 27.62 & 19.75 & 19.30  & 30.19 & 34.81 & 4.78  \\
\hline
$\mathcal{S}$ ($B$: $m_{Z'}=800$~GeV) & 3.30 & 2.34 & 2.33  & 4.88 & 6.28 & 0.48  \\
\hline\hline
$\mathcal{S'}$ ($A$: $m_{Z'}=400$~GeV) & 12.03 & 9.35 & 9.73  & 13.98 & 16.30 & 2.70  \\
\hline
$\mathcal{S'}$ ($B$: $m_{Z'}=800$~GeV) & 2.36 & 1.85 & 2.17  & 4.05 & 5.55 & 0.47  \\
\hline
\end{tabular}
\caption{\label{tab:lepton_charge}
 Signal significance for the benchmark point $A$ and $B$ for each
 positively and negatively charged lepton in the final state.
}
\end{table}

The charge selection can efficiently distinguish between the top--charm
and top--up DM models. 
For the top--up model, the significance increases (decreases) remarkably
by requiring only positively (negatively) charged leptons.
The predominance of positively charged configurations in the top--up
model implies that the significance of the analysis targeted to a
positively charge lepton is even larger than the one without the charge
identification. 
For the top--charm DM model, on the other hand, the significance for the
positively charged case and the negatively charged case are essentially
equivalent, and they are both smaller than the combined one.
In short,
\begin{align}
 &\mathcal{S}_{\ell^++\ell^-} > 
 \mathcal{S}_{\ell^+} \simeq \mathcal{S}_{ \ell^-}
 &&\text{for the top--charm DM model}\,, \\
 &\mathcal{S}_{\ell^+} > 
 \mathcal{S}_{\ell^++\ell^-} \gg \mathcal{S}_{ \ell^-}
 &&\text{for the top--up DM model}\,,
\end{align}
and we conclude that the lepton charge identification provides an
efficient technique to distinguish between the top--charm and top--up DM
models in the monotop signature.

\paragraph{Charm-jet tagging:}

The second strategy that we investigate to distinguish between the
top--charm and the top--up models makes use of a charm tagging for an
extra jet in SR2.

Such a charm-tagging algorithm has been recently released by
ATLAS~\cite{ATL-PHYS-PUB-2015-001} and it exploits the properties of
displaced tracks, reconstructed secondary vertices and soft leptons
inside jets. 
In our analysis, we assume a constant tagging efficiency for simplicity,
without any dependence on $p_{T}$ and $\eta$ of the jets. 
This is accurate enough for a first estimation of the effect of charm
tagging as the $p_T$ and $\eta$ dependence is quite
mild~\cite{ATL-PHYS-PUB-2015-001}. 
Based on the needs of a specific analysis, different working points can
be chosen to select the desired charm-jet tagging efficiency and to
either improve the rejection of light-flavour jets or bottom-quark
jets. 

To select the charm-flavour jet with a low mistag rate for light
($u,d,s,g$) jets, we employ a tight $c$-tagging working point.
Taking inspiration from the efficiency performance reported by the
ATLAS Collaboration~\cite{ATL-PHYS-PUB-2015-001}, we assume an overall
$c$-tagging efficiency of 20\,\% for $c$-flavour jets, 1\,\% for
$u,d,s$-flavour and gluon jets, and 15\,\% for $b$-flavour jets.
We note that the mistag rate is extremely low, but we are compromised by
the rather low $c$-tagging efficiency.
Ideally, if the charm tagger would have a 100\,\% efficiency, 
the signal cross section would be suppressed by roughly a factor of a
hundred for the top--up model, while by about a factor of three for the
top--charm model, as seen in Fig.~\ref{fig:xsec}.

\begin{table}
\center
\begin{tabular}{|l|r|r|r||r|r|r|}
\cline{2-7}
   \multicolumn{1}{c}{}  & \multicolumn{3}{|c||}{Before $c$-tagging} 
 & \multicolumn{3}{c|}{After $c$-tagging}   \\
 \cline{2-7}  
 \multicolumn{1}{c|}{} & SR1 & SR2 & SR1+2 & SR1 & SR2 & SR1+2   \\
\hline
$\mathcal{S}$ (top--up; $A$) & 24.78 & 18.10 & 30.19 & 24.78 & 1.19 &
 24.44  \\
 \hline
 $\mathcal{S}$  (top--charm; $A$) & 23.10 & 16.15 & 27.62 & 23.10 & 2.30 & 23.02  \\
 \hline
 \hline
$\mathcal{S}'$ (top--up; $A$) & 12.64 & 9.46 & 13.98 & 12.64 & 1.52 & 12.72  \\
\hline
$\mathcal{S}'$  (top--charm; $A$) & 10.19 & 8.99 & 12.03 & 10.19 & 2.74 &10.49  \\
\hline
\end{tabular}
\caption{\label{tab:CTag_Compare_ctut}
 Signal significance for the benchmark point $A$ in the both top--up and
 top--charm models before and after the charm-tagging requirement.
}
\end{table} 

Considering the same cuts of the previous section, we require the second
jet of the SR2 to be a charm-tagged jet, and compute the significance
for the benchmark point $A\,(m_{Z'}=400~{\rm GeV})$ for the both top--up 
and top--charm models.   
Table~\ref{tab:CTag_Compare_ctut} shows that after $c$-tagging the significance of the SR2
for the top--charm model becomes twice larger than that for the top--up
model although the signal significance itself suffers a sharp drop for
the both models.

Even though this technique is probably much less efficient than the
lepton charge asymmetry, it represents nevertheless an alternative
strategy to distinguish between the top--up and top--charm DM models in
the monotop signature.
More detailed investigations of charm-tagging
techniques could result in better performances in the search for DM
described in this paper.

%%%%%%%%%%%%%%%%%%%%
\subsection{Canonical dark matter searches}\label{sec:cDMsearch}
%%%%%%%%%%%%%%%%%%%%

A stable new particle that interacts predominantly with a top--charm or
a top--up quark pair has interesting implications for DM phenomenology,
in particular for relic density and canonical DM searches such as direct
and indirect detection experiments.
In the following we discuss in detail these considerations, devoting
special attention to a tentative explanation of the galactic centre
excess in terms of flavour-changing DM. 

%%%
\paragraph{Relic density:}
%%%

As mentioned in Sec.~\ref{sec:models}, we focus on the parameter
space where $m_{Z'}$ is parametrically the highest scale
of the theory.
In this case, the typical energy scale of the thermal freeze-out process
is smaller than the mass of the mediator $m_{Z'}$.
We then expect that the description of the dynamics in terms of a
simplified model or an EFT makes no practical difference.%
\footnote{Apart from parametrically separating $m_{Z'}$ from the other
scales of the theory, one has to ensure that the width of $Z'$ is not
too large, i.e. $\Gamma_{Z'}< m_{Z'}$.}
Indeed, for the parameter points of interest, we checked that the
results in the simplified model~\eqref{ZprimeLag} and in the EFT
description~\eqref{eftLag} agree very well, by using 
{\sc MicroOMEGAs}~\cite{Belanger:2006is,Belanger:2008sj}
and {\sc MadDM}~\cite{Backovic:2013dpa,Backovic:2015cra}.

\begin{figure}
\center
 \subfigure[
 We assume the observed DM relic density in all of the parameter plane. 
 The area within the dashed boundary shows the parameter region that can
 fit the galactic centre excess. 
The grey area is excluded by FERMI. 
]{
            \label{DM_1}
            \includegraphics[width=0.47\textwidth]{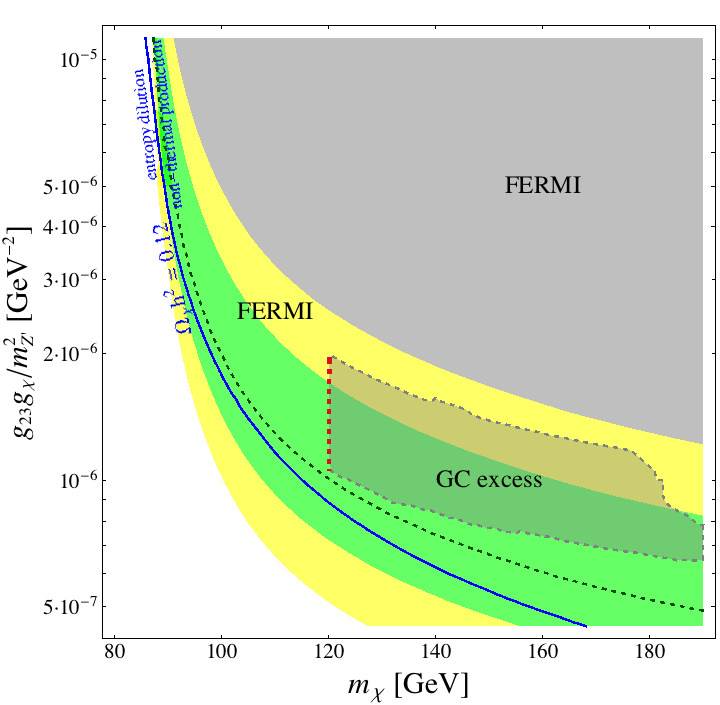}
        }
        \hfill
        \subfigure[
 We do not make such assumption as in Fig.~\ref{DM_1}, and 
hence the FERMI limit is rescaled according to the relic density of the
 model. 
 The dark grey area is excluded by the DM overabundance.
]{
            \label{DM_2}
            \includegraphics[width=0.47\textwidth]{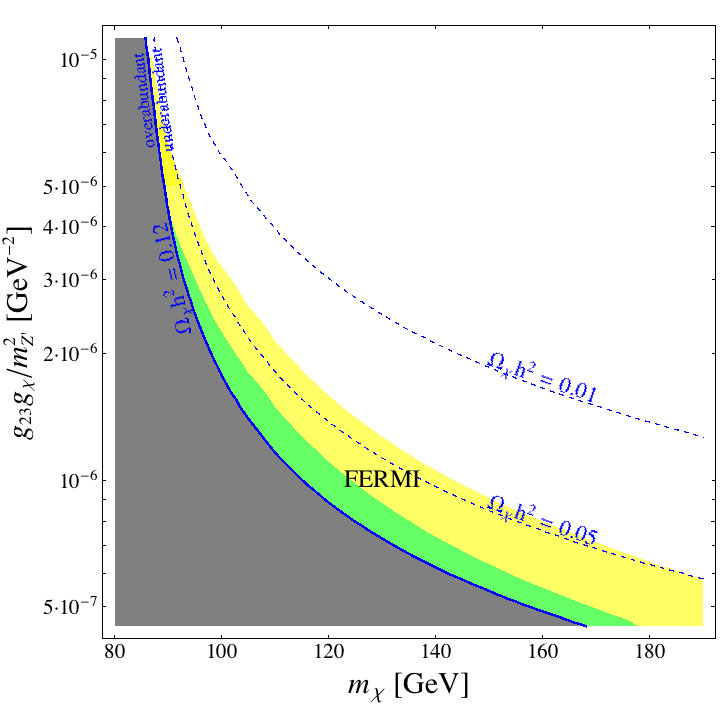}
        }\\ 
\caption{
 Relic density contours (blue lines), limits from FERMI (dashed green
 line with green ($1\sigma$) and yellow ($2\sigma$) expected uncertainty
 bands),  
 and the parameter region that fits the
 galactic centre excess (light grey region) for the top--charm
 flavour-changing DM model. 
     }%
     \label{fig:DM}
\end{figure}

In the simplified model, the annihilation of the DM candidate to the
quark pair during the thermal freeze-out occurs exclusively via
$s$-channel mediation of a $Z'$ boson, as shown in Eq.~\eqref{annihilation}.
The relic density of $\chi$ can be equal to the observed relic density
of DM, $\Omega_\chi h^2=0.12$~\cite{Ade:2013zuv} for reasonable values
of the effective coefficient $g_{i3}g_{\chi}/m_{Z'}^2$. This is
illustrated for the top--charm model in Fig.~\ref{fig:DM}, while the
corresponding plot for the top--up model looks practically the same, as
the mass difference between the charm and up quarks has little effect on both the calculation of the relic density and the photon fluxes of the annihilation products.

In Fig.~\ref{DM_1}, the proper relic density is depicted by the blue contour, however it is assumed that the relic density of $\chi$ is equal to the observed DM relic density in all of the parameter space. 
For $m_\chi<m_t/2$ the annihilation to $t$--$c$ or $t$--$u$ quark pair
is kinematically forbidden and the correct relic density is achieved if
we invoke extra dynamics, e.g. an entropy dilution mechanism.
For values of the parameter space to the right of the
$\Omega_\chi h^2=0.12$ line, $\chi$ would annihilate too much and the
proper relic density is achieved if we assume non-thermal production. 
In Fig.~\ref{DM_2}, on the other hand, we do not make such assumptions,
so that the region of the parameter space with overabundant DM is
excluded while in the region where DM is underabundant we provide
indicative relic density contours. 
 
%%%
\paragraph{Indirect searches:}
%%%

The latest FERMI data on photon fluxes from dwarf spheroidal galaxies of
the Milky Way~\cite{Ackermann:2015zua} provide strong constraints on the
parameter space of the model, as illustrated in Fig.~\ref{fig:DM}.
Although the published results do not include annihilation to
flavour-violating pairs such as $t$-$c/u$, the photon flux is very
similar to that of $b\bar b$ for the DM mass range under
focus~\cite{Cirelli:2010xx}. 
Therefore, the limit on DM annihilating to $t$-$c/u$ is similar to that
of a bottom-quark pair.
The dashed green line shows the observed limit while the green and
yellow bands depict the $1\sigma$ and $2\sigma$ uncertainties of the
expected sensitivity.
We notice that the observed limit is nearly probing the thermal relic
density line of the model. 
In Fig.~\ref{DM_2} we show the FERMI limits rescaled by
$(\Omega_\chi h^2 / 0.12)^2$ to account for the varying relic density. 
We note that, while in Fig.~\ref{DM_1} FERMI excludes all the large
$m_\chi$ region (because of the increased annihilation), in
Fig.~\ref{DM_2} the decreased relic density dominates over the increased
cross section so that FERMI turns out to be sensitive only up to the
$\Omega_\chi h^2\simeq 0.05$ line. 

The latest results from the searches for antiprotons by the AMS-02
experiment~\cite{ams02} can also be used to set limits on DM
annihilation in the centre of our galaxy.
The reach of these limits depends on the uncertainties of the
astrophysical background and the propagation of the antiprotons in the
galaxy~\cite{Giesen:2015ufa,Jin:2015sqa,Kappl:2015bqa}, and under
reasonable assumptions the limit on DM annihilating to $b\bar b$ is
equivalent or even stronger than the one obtained from FERMI.
Due to the relatively larger uncertainties, we have not used these
limits in this work, however, it would be interesting to see how much
further they can constrain the models discussed here. 

We also report on the excess of gamma rays from the galactic centre that
has been observed~\cite{Hooper:2010mq} and updated~\cite{murgia} by the
FERMI telescope. Initial proposals that fit the photon profile of the excess included DM annihilating to $b\bar b$ pairs (see eg \cite{Berlin:2014tja,Alves:2014yha} for phenomenological analyses). The top--charm model leads to a simlar photon flux, and hence can fit
the excess~\cite{Rajaraman:2015xka}.
In Fig.~\ref{DM_1} we show, based on the
results of~\cite{Rajaraman:2015xka}, the region of the parameter space
of the model that can fit the galactic centre excess.
The fit to the excess corresponds to DM that is slightly
over-annihilating, so that a non-thermal production mechanism is
required to ensure the observed relic density. 
The abrupt stop at $m_\chi=120$~GeV is an artifact of extracting photon fluxes from PPPC~\cite{Cirelli:2010xx}; in
principle one expects the fit region to expand to lower DM masses.%
\footnote{The photon flux for annihilation to a top-quark pair given in
PPPC starts from $m_\chi = 180$~GeV. When recasting the top-pair and
charm-pair fluxes to get the flux of the top--charm pair, this
translates into a minimum value of $m_\chi = 120$~GeV.}
We also note that, since the photon flux from an up quark is practically the same to that from a charm quark~\cite{Cirelli:2010xx}, the top--up model fits the galactic centre excess, too.

%%%
\paragraph{Direct searches:}
%%%

The DM candidate $\chi$ 
in the top flavour-changing model 
has radiatively induced flavour--conserving interactions
with quarks so that, in principle, direct search experiments can be
relevant.\footnote{There are also effective interactions with gluons which are however higher dimensional and two--loops suppressed, therefore they are negligible in our simplified model.}

The reactor response depends on the strength of the interaction between the DM candidate and the reactor nucleus, typically described in terms of non-relativistic effective operators that in general depend on the momentum exchange between the two particles, their relative velocity and their spins. These, in turn, are described in terms of effective $\chi$-$N$ operators, themselves described in terms of a set of effective interactions between $\chi$ and the quarks or gluons.

In describing the scattering of the DM candidate against nuclei for
low-energy experiments, we can take the limit of small relative velocity
and momentum transfer.
In this limit and neglecting higher derivative operators, the scalar
($\ov{\chi}\chi\, \ov{q}q$) and vector 
($\ov{\chi}\gamma^\mu\chi\,\ov{q}\gamma_\mu q$) DM--quark interactions contribute to the scattering
cross section that does not depend on the spin of the colliding
particles, 
the axial-vector ($\ov{\chi}\gamma^\mu\gamma^5\chi\,
\ov{q}\gamma_\mu \gamma^5q$) and tensor ($\ov{\chi}\sigma^{\mu\nu}\chi\,
\ov{q}\sigma_{\mu\nu} q$) interactions contribute to spin-dependent
scattering, while all the other effective operators can be neglected.

For the top--up model the interaction of DM with a nucleus is achieved
via box diagrams with two $Z'$ bosons, one top quark and one DM field
running in the loop, while for the top--charm model the connection with
the valence quarks of the nucleon requires a second loop. We have
calculated them in the zero momentum transfer limit for the top--up
model and matched the Wilson coefficients to the simplified model computation. Regarding the spin-independent cross sections, in the limit of a massless up-quark, the contribution to the scalar operator vanishes, while
the finite contribution to the vector operator $c_V\ov{\chi}\gamma^\mu\chi\,\ov{q}\gamma_\mu q$ is given by the following Wilson coefficient:
\be
c_V=\!6\,g_\chi^2\,g_{13}^2\hspace{-0.2cm}\int_0^1\!\!\!\!\! dx\!\! \int_0^{1-x}\hspace{-0.73cm} dy\ x \Big[{3m_\chi^2\over 2m_{Z'}^2}z\!\left(\! I_{l_1^2}^{n=4}\!-\!I_{l_2^2}^{n=4} \!\right)\!-m_{\chi}^2 z\!\left( I_{0_1}^{n=4}\!-\!I_{0_2}^{n=4} \right)\!+2\!\!\!\int_0^{1-x-y}\hspace{-1.1cm} dz\  z\, m_\chi^2(I_{l_1^2}^{n=5}-\!I_{l_2^2}^{n=5})\! \Big]\,,
\nn
\ee
where $z=1-x-y$ when it is not the variable of integration and 
\bea
&&I_{0_i}^{n}=\int {d^4 l_i\over (2\pi)^4}{1\over (l_i^2-\Delta_i)^n}=2i{(-1)^{n}\over 16\pi^2}{\Gamma(n-2)\over \Gamma(n)}{1\over \Delta_i^{n-2}}\,,\nn
\\
&&I_{l^2_i}^{n}=\int {d^4 l_i\over (2\pi)^4}{l_i^2\over (l_i^2-\Delta_i)^n}=i{(-1)^{n-1}\over 16\pi^2}{\Gamma(n-3)\over \Gamma(n)}{1\over \Delta_i^{n-3}}\,,
\eea
with
\bea
&&l_1=q+yk-zp\,,\quad \nn \Delta_1=(yk-zp)^2+xm_{Z'}^2-y(m_u^2-m_t^2)\,,
\\
&&l_2=q+yk+zp\,,\quad \nn \Delta_2=(yk+zp)^2+xm_{Z'}^2-y(m_u^2-m_t^2) \,,
\eea
where $p$, $k$ and $q$ are the DM, the quark and the loop momentum respectively. For reasonable values of the model parameters, the size of the Wilson coefficient turns out to be $\lesssim 10^{-50}\,\textrm{cm}^2$, orders of magnitude smaller than current observational sensitivities, due to the cancellation among the box diagrams. This is true also for the spin-dependent cross section, where the experimental constraints are less strong. Therefore, the top--up DM model is not constrained by direct search experiments and consequently, neither is the top--charm model.

%%%%%%%%%%%%%%%%%%%%%%%%%%%%
\subsection{Complementarity between the LHC and non-collider
  experiments}
\label{sec:combo}
%%%%%%%%%%%%%%%%%%%%%%%%%%%%

In the previous subsections we have studied the LHC and non-collider
phenomenology separately for the top flavour-changing simplified DM
model.
Here we combine the two analyses to provide a complete picture of the
experimental reach on the parameter space of the model and the
complementarity among different DM search experiments.

\begin{figure}
\center
 \includegraphics[width=0.475\textwidth]{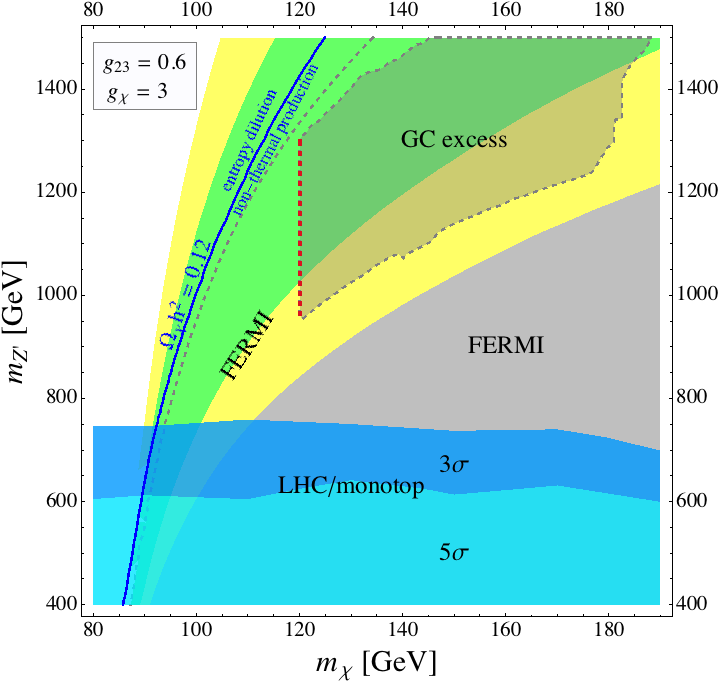}\quad
 \includegraphics[width=0.475\textwidth]{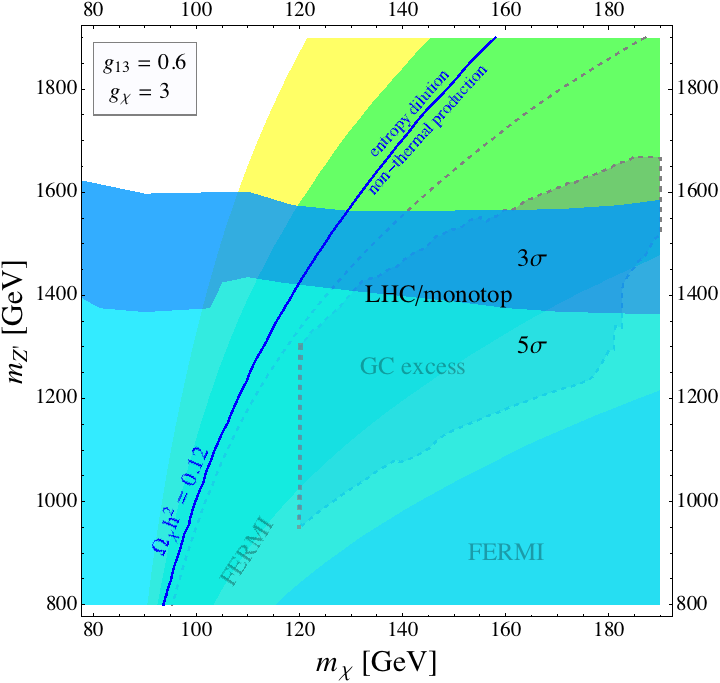}
\caption{\label{fig:SUM}
 Aggregated figures of the relic density, FERMI limits and LHC reach for
 the DM candidate of the top--charm (left) and
 top--up (right) flavour-changing models in the DM--mediator mass plane,
 where we assume the observed DM relic density in all of the parameter
 plane. 
The grey area with dashed boundary shows the parameter region that can
 fit the galactic centre excess. The dashed green line is the observed FERMI
 limit, while the green and yellow bands correspond to $1\sigma$ and $2\sigma$
 uncertainties on the expected limit. The dark and light blue bands
 depict the $3\sigma$ and $5\sigma$ reach of monotop searches at LHC-13TeV with 100 fb$^{-1}$.
}
\end{figure}

The results are summarised in Fig.~\ref{fig:SUM}, where we show the
prospects from LHC-13TeV and the FERMI constraints together with the
region of the parameter space that fits the tentative galactic centre
excess for both the top--charm and the top--up models.   
In these plots the FERMI limits are obtained assuming that the relic
density of the DM is equal to the observed one in all the parameter
plane, allowing for other mechanisms than thermal production, same as in
Fig.~\ref{DM_1}. 
If we would instead assume only thermal production for the DM candidate,
the observed relic abundance is obtained only along the blue lines and
on the rest of the  parameter space the bounds from FERMI are much
weaker (see Fig.~\ref{DM_2} and discussion there).  
The LHC-13TeV reach, on the other hand, does not depend on these
assumptions.
This is already a basic difference between the limits derived from
colliders and from indirect detection. 

The LHC reach depends almost exclusively on the mediator mass, which
sets the size of the cross section (for fixed couplings).
On the other hand, the reach of indirect detection experiments depends
also on the DM mass, which affects the efficiency of the DM
annihilation.  
This implies that the LHC and indirect DM experiments can probe
different regions of the parameter space of the model. 

Another interesting point we observe is that, by analysing the two plots
in Fig.~\ref{fig:SUM}, a combined interpretation of the top
flavour-changing DM models at LHC-13TeV and in indirect DM searches
reveals different features between the top--charm and the top--up DM
model.  

In the top--charm DM model (Fig.~\ref{fig:SUM}(left)), the FERMI exclusion
covers most of the parameter space that can be probed by LHC-13TeV.  
However, the blue line where the dark matter abundance is obtained via
usual thermal production is not constrained by FERMI, and instead it
will be probed by LHC-13TeV for a DM mass around 90~GeV. 
The region capable of explaining the galactic center excess,
characterized by the mediator heavier than a TeV, lies beyond the reach
of LHC-13TeV. 

The top--up DM model (Fig.~\ref{fig:SUM}(right)) presents the same limit from
indirect detection as in the top--charm DM model 
but has a much larger reach at LHC-13TeV.
The LHC-13TeV will be able to probe the thermal relic DM line up to a
mass of around 130~GeV, and to cover almost completely the region
capable of accommodating the galactic center excess.

Even though in the figures we have fixed the coupling as
$(g_\chi,g_{i3})=(3,0.6)$, the previous discussion is robust under
modifications of the $g_{i3}$ coupling as long as the invisible decay
of the $Z'$ remains dominant. 
This is due to the fact that the monotop signature scales as 
$g_{i3}^2/m_{Z'}^4\times B(Z'\to\chi\bar\chi)$ and the DM annihilation 
scale as $g_{i3}^2 g_{\chi}^2/m_{Z'}^4$.
Hence, reducing the coupling $g_{i3}$ (keeping the $Z'$ invisible decay
as the dominant one) will shift down by the same amount both the region
capable of fitting the galactic centre excess as well as the $3\sigma$
and $5\sigma$ discovery lines of LHC-13TeV, and thus it will not affect
qualitatively our conclusions.
In this perspective, one can argue that the monotop signature at the LHC
and the canonical DM searches in this simplified model allow for a
straightforward comparison, because of their similar scaling with the
couplings.

%%%%%%%%%%%%%%%%%%%%%%%%%%%%
\section{Conclusions and discussions}\label{sec:summary}
%%%%%%%%%%%%%%%%%%%%%%%%%%%%

In this work we have studied the phenomenology of a simplified model of
DM with flavour-changing interactions.
Given the strong constraints on flavour-changing interactions of the
down-quark sector from low-energy experiments, we focused on DM
interacting with a right-handed top--up or top--charm pair via a neutral
vector mediator $Z'$. 
The simplified model is parametrised by the mass of the DM candidate,
the mass of the mediator and the couplings of the $Z'$ to the DM and the
quark pair.
Depending on these parameters, the model provides rich signatures at
colliders as well as at non-collider experiments, as summarised in
Table~\ref{tab:sig} and described in Sec.~\ref{sec:modelsig}.

We focused on the top--charm flavour-changing DM model whose most
relevant signature at the LHC is a single top quark plus missing energy,
i.e. a monotop final state.
For our benchmark point $g_\chi=3$ and $g_{23}=0.6$, the limit from
LHC-8TeV is approximately $m_{Z'}\gtrsim 400$~GeV.  
For the prospects of LHC-13TeV with 100~fb$^{-1}$, we find that, for the
same couplings, the $3\sigma$ ($5\sigma$) reach can go up to 
$m_{Z'}\sim 760\ (640)$~GeV, roughly independent of the DM mass. 
We then discussed how to distinguish the top--charm DM model from the
top--up one in the monotop signatures by making use of lepton charge
determination and by employing a charm-tagging technique.

For non-collider DM signatures,
we showed that the DM candidates with top flavour-changing interactions
can be thermal relics for reasonable values of couplings and for a mass
of the order of electroweak scale.  
We found that direct searches do not pose bounds on the simplified
models under study, due to the cancellation of the box diagrams involved
in the scattering of DM against nuclei.
On the other hand, indirect searches pose strong bounds.
We used the results from FERMI on photon fluxes from dwarf spheroidal
galaxies to constrain the parameter space of the models and identified
the part of the parameter space that fits the galactic centre excess.
Since the photon fluxes of the top--up and top--charm models are
practically same, the both models fit the excess equally well. 

Finally, combining the LHC and non-collider analyses, we showed the
complementarity among the different DM search experiments in probing the
parameter space of the model and how the combination of these analyses
will be able to distinguish between the top--charm and top--up
flavour-changing DM models.\\  

Before closing, we would like to comment on the UV completion of the model for what concerns the origin of the flavour-changing couplings and the extra degrees of freedom needed in order to make the model anomaly free. Considerations regarding the UV completion of simplified models with $Z'$ bosons have been recently discussed in \cite{Kahlhoefer:2015bea}.

One way to build the flavour-changing terms is to impose different
charges under the $U(1)'$ gauge symmetry for each generation of
quarks. After switching to the mass eigenstate basis, the coupling of
the quarks to $Z'$ can be written as $g'Q_{ij}\ov{u}_{R}^i\gamma^\mu
u_{R}^jZ'_\mu$, where $Q_{ij}=Q'_kV^{R\dag}_{ik}V^R_{kj}$ and $V^{R,L} $
are the unitary matrices that diagonalise the quark mass matrix and
$Q'_i$ is the gauge charge of the quark of the $i^{\textrm{th}}$
generation under the $U(1)'$.
In our model, we choose $g_{i3}=g'Q_{i3}$ where $i=1,2$ and set all the rest to zero. 

Since we do not want to charge the left-handed quarks under $U(1)'$, there
are two ways to render the SM Yukawa couplings gauge invariant. One way
is to charge the Higgs boson. Since every generation has different
charge $Q_i'$, we would need to introduce a different Higgs boson (with
charge $-Q_i'$) for every generation. This leads to theories with extra
Higgs doublets, discussed in the past in the context of the
forward-backward asymmetry~\cite{Ko:2011vd}. The second way is to use
the Froggatt--Nielsen (FN) mechanism, i.e. to interpret the Yukawa coupling as the expectation value of a dynamical scalar field $\phi$ that is charged under $U(1)'$. Either way, the construction of the flavour-changing $Z'$ coupling requires extra scalars that are charged under $U(1)'$, either extra Higgs doublets or an extra FN type scalar. In our work we focused on model-independent aspects of the top flavour-changing DM model by assuming that these extra states are heavy enough so as not to play a role in LHC or DM detection experiments.

As for the second point, the model per se is anomalous. Charging only the right-handed quarks  under the $U(1)'$ introduces gauge anomalies from triangle diagrams that involve the $Z'$ and SM gauge bosons. Phenomenological and theoretical aspects of anomalous $U(1)'$ extensions of the SM have been extensively discussed, see~\cite{Anastasopoulos:2006cz,Mambrini:2011dw,Dudas:2013sia} and references therein. In order to cancel the anomalies, new chiral fermions $\psi_{L,R}$ need to be added that are also charged under $U(1)'$ and the SM gauge groups. These chiral fermions get their mass by the spontaneous breaking of the $U(1)'$ gauge symmetry, via Yukawa interactions of type $y\,\varphi\, \ov{\psi}_{L}\psi_R$, so that $m_\psi \sim y\, v_\varphi$ while $m_{Z'}= g'v_\varphi/2$.
Therefore, for moderate $g_{i3}$ and for typical $U(1)'$ charge assignments we expect that the mass of these fermions is not much heavier than $m_{Z'}$,
however leaving enough room to consider these extra states beyond the reach of LHC for large Yukawa couplings.
Indeed, in our phenomenological analysis we take $g_{i3}=0.6$ which
suggests that the extra fermion masses can be easily larger than the TeV
scale, which is beyond the current bound on heavy quarks of 950
(782)~GeV from the ATLAS~\cite{Aad:2015kqa} (CMS~\cite{Chatrchyan:2013uxa}). Furthermore, we estimated that for $m_{Z'}\gtrsim 400\,$GeV there are no bounds from current direct search experiments coming from the effective $Z/Z'$ kinetic mixing \cite{Mambrini:2011dw} or from the effective $Z'$-$g$-$g$ coupling.

Summarizing, in our work we have focused on model-independent aspects of the top flavour-changing DM model and neglected extra states related to possible UV completions 
by assuming that they are heavy enough to not affect the phenomenology significantly. 
It would be interesting to study the signatures of these states and obtain combined constraints by associating it with the analyses we performed.

%%%%%%%%%%%%%%
\section*{Acknowledgements}
%%%%%%%%%%%%%%

The authors would like to thank fruitful discussions with
P.~Anastasopoulos, L.~Calibbi, E.~Dudas and B.~Zaldivar. 
This work is supported in part by Vrije Universiteit Brussel
through the Strategic Research Program `High-Energy Physics', 
and
in part by the Belgian Federal Science Policy Office through the
Inter-university Attraction Pole P7/37.
A.M. and P.T. are also supported in part by FWO-Vlaanderen through
project G011410N. 
A.M. aknowledges the Pegasus FWO postdoctoral Fellowship.
K.M. is supported by the ``Theory-LHC-France Initiative'' of CNRS
(INP/IN2P3). 
S.M. is a Aspirant van het Fonds Wetenschappelijk Onderzoek - Vlaanderen.

%\appendix

\bibliography{bibfvtopdm}
\bibliographystyle{JHEP}

\end{document}